\documentclass[letterpaper]{article} 
\usepackage[submission]{aaai2026}  
\usepackage{times}  
\usepackage{helvet}  
\usepackage{courier}  
\usepackage[hyphens]{url}  
\usepackage{graphicx} 
\usepackage{natbib}  
\usepackage{amsfonts}
\usepackage{caption} 
\usepackage{booktabs}
\usepackage{multirow}
\usepackage{algorithm}
\usepackage{algorithmic}
\usepackage[activate={true,nocompatibility},tracking=true,spacing=true,kerning=true,final]{microtype}
\usepackage[english]{babel}
\usepackage{microtype}
\usepackage{amsmath}
\usepackage{newfloat}
\usepackage{listings}
\usepackage{times}
\usepackage{helvet}
\usepackage{courier}
\usepackage{xcolor}
\DeclareCaptionStyle{ruled}{labelfont=normalfont,labelsep=colon,strut=off} 
\floatstyle{ruled}
\newfloat{listing}{tb}{lst}{}
\floatname{listing}{Listing}
\title{From Stimuli to Minds: Enhancing Psychological Reasoning in LLMs\\via Bilateral Reinforcement Learning}
\author {
    Yichao Feng\equalcontrib,
    Haoran Luo\equalcontrib,
    Lang Feng,
    Shuai Zhao,
    Anh Tuan Luu
}
\affiliations{
    Nanyang Technological University\\
    College of Computing and Data Science\\
    \{yichao.feng, lang.feng, shuai.zhao, anhtuan.luu\}@ntu.edu.sg, haoran.luo@ieee.org
}

\usepackage{bibentry}

\begin{document}
\maketitle
\begin{abstract}
Large Language Models show promise in emotion understanding, social reasoning, and empathy, yet they struggle with psychologically grounded tasks that require inferring implicit mental states in context-rich, ambiguous settings. These limitations arise from the absence of theory-aligned supervision and the difficulty of capturing nuanced mental processes in real-world narratives. To address this gap, we leverage expert-labeled, psychologically rich scenarios and propose a trajectory-aware reinforcement learning framework that explicitly imitates expert psychological thought patterns. By integrating real-world stimuli with structured reasoning guidance, our approach enables compact models to internalize social-cognitive principles, perform nuanced psychological inference, and support continual self-improvement. Experiments across multiple benchmarks demonstrate that our models achieve expert-level interpretive capabilities, exhibiting strong out-of-distribution generalization across diverse psychologically tasks. Our code is publicly available at \url{https://github.com/Githubuseryf/Stimuli2Minds}.
\end{abstract}
\section{Introduction}

\label{s1}

Large Language Models (LLMs) show strong generalization across diverse language tasks \cite{wu2025sailing,zhang2025timemaster,feng2025aspect}. Their emerging potential in psychological domains, such as emotion understanding \cite{kovacevic2024multimodal}, social reasoning \cite{leng2023llm}, and empathy recognition \cite{sorin2024large} has drawn growing research interest. These tasks require not only linguistic comprehension but also nuanced inference of implicit mental states and emotional cues, often without explicit supervision or clearly defined ground truths. Unlike conventional language tasks, they involve rich psychological stimuli embedded in ambiguous, socially grounded scenarios shaped by diverse cultural norms, interpersonal dynamics, and lived experiences. Although LLMs show partial sensitivity to such cues, performance on psychologically grounded tasks remains limited from human-level competence \cite{ke2025exploring}.
\begin{figure}[t]
    \centering
    \includegraphics[width=\linewidth]{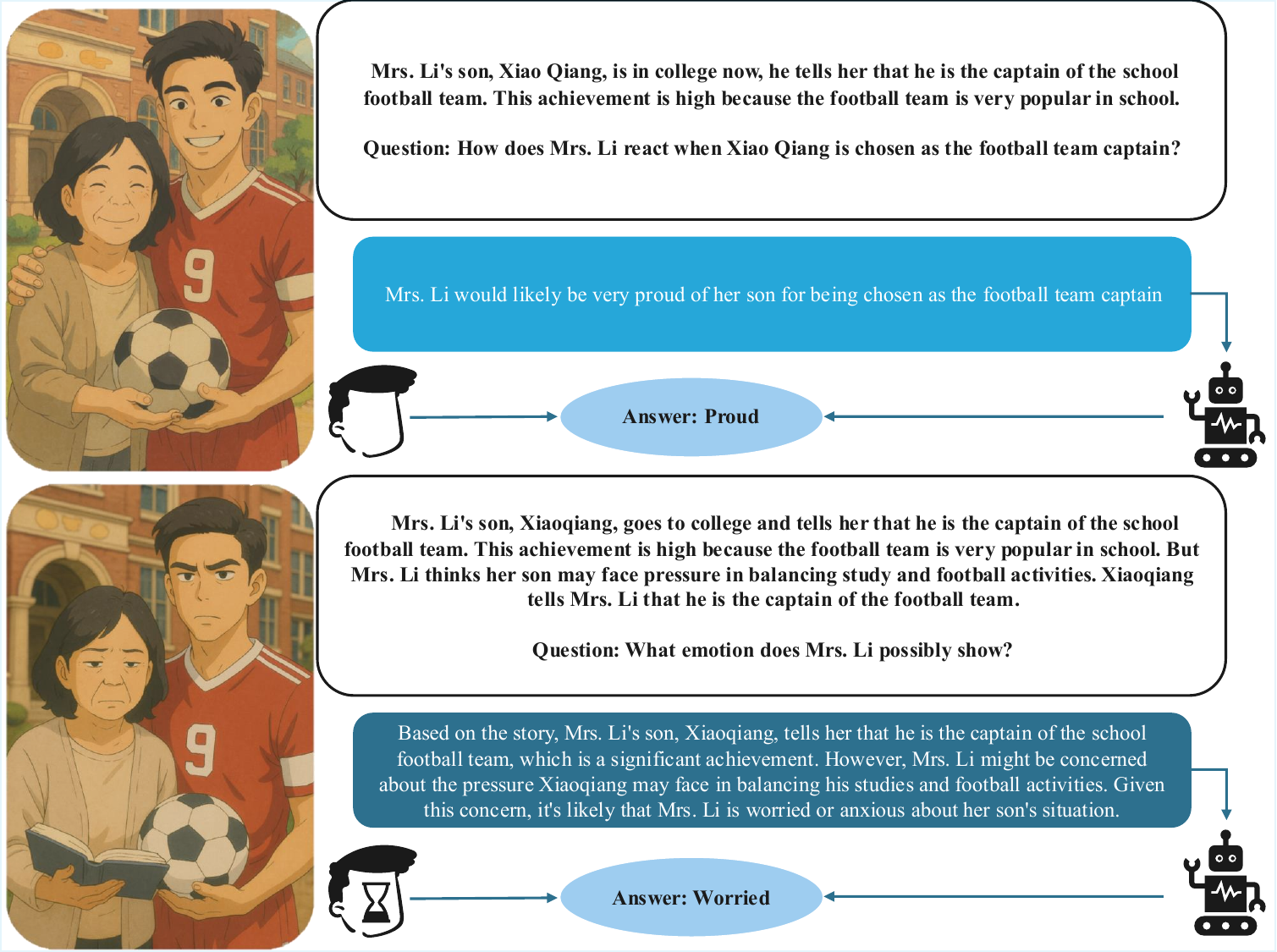}
    \caption{A sample ToMbench question presents two types of psychological stimuli pairs for demonstration.}
    \label{fig:think_think}
\end{figure}
\begin{figure*}[!t]
    \centering
    \includegraphics[width=\linewidth]{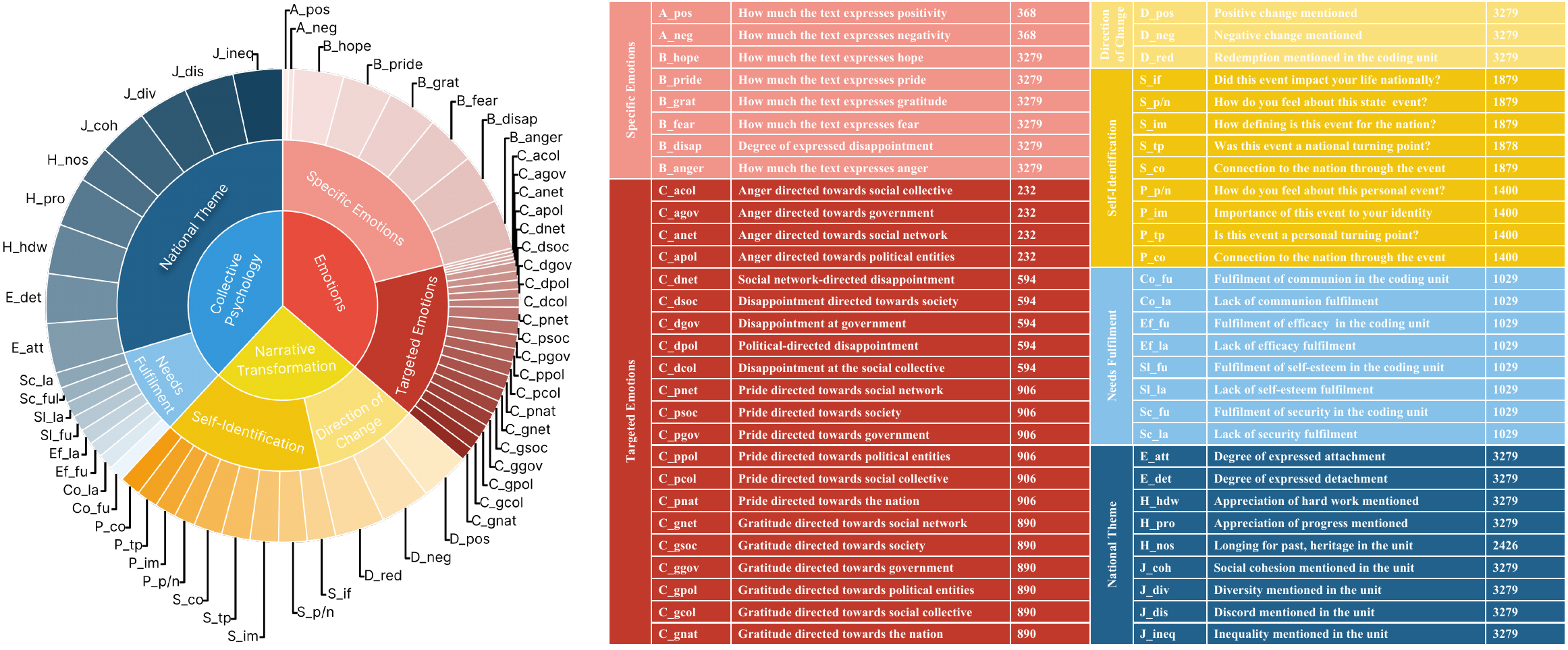}
    \caption{The figure summarizes key psychological parameters across our datasets: emotional variables, narrative transformations, and collective factors, comprising 35,084, 24,831, and 36,890 QA instances, each offering insights into human narratives.}
    \label{fig:StimuliQA_structure}
\end{figure*}

To evaluate LLMs’ capacity for cognitive reasoning, several benchmarks have emerged. \textbf{ToMbench} provides a structured multiple-choice framework for assessing Theory of Mind (ToM) across eight task types and 31 social-cognitive capabilities \cite{chen2024tombench}, and shows that even advanced models like GPT-4o \cite{achiam2023gpt} still lag behind human performance. Other benchmarks such as Psychobench \cite{li2024quantifying} and CogBench \cite{coda2024cogbench} further explore emotion inference and pragmatic reasoning. These evaluations mark a growing effort to systematize measurement of social-cognitive skills in LLMs. While LLMs display partial sensitivity to mental states, their reasoning remains fragmented, often relying on surface cues rather than robust mental models and struggle to generalize in complex scenarios, falling short of human-like social understanding.

Despite recent advances, three key challenges remain.
\textbf{(1)  High quality data scarcity:} Benchmarks like ToMbench are small in scale \cite{wu-etal-2023-hi}, and many datasets use LLM-generated content \cite{hu2024psycollm}, limiting their value for tuning psychological reasoning \cite{long-etal-2024-llms}.
\textbf{(2) Reasoning mismatch across tasks:} Psychological tasks differ in cognitive demands. Theories as Epstein’s \cite{epstein1998cognitive} and Fuzzy-Trace Theory \cite{reyna1998fuzzy} separate fast, intuitive reasoning from slow, analytical thought. Uniform strategies may hurt performance on intuition-driven tasks \cite{ji2025difficulty}. As an example illustrated in Figure~\ref{fig:think_think}. \textbf{(3) Poor generalization in small models:} Compact LLMs tend to overfit and struggle to generalize. Many depend on LLM-generated labels or costly prompting \cite{ijcai2024p719}.

To address these limitations, we make three contributions. 
\textbf{First}, we construct a large-scale dataset, \textbf{StimuliQA}, grounded in professional psychological theory and real-world interviews. 
It features over 3{,}000 annotated stimuli with 58 psychological variables, which are subsequently converted into question–answer (QA) pairs, as detailed in Figure~\ref{fig:StimuliQA_structure}. 
\textbf{Second}, we introduce \textbf{Psy-Interpreter}, a reinforcement learning (RL) framework inspired by dual-system psychological theories. 
It incorporates a trajectory cache and bilateral reasoning (BR) to foster expert-like psychological analysis across diverse tasks. 
\textbf{Third}, we demonstrate that compact models, when trained with the Psy-Interpreter framework and rationale-augmented supervision, can rival much larger systems across multiple benchmarks, highlighting the effectiveness of structured RL and reasoning-aware supervision in enhancing efficiency and generalization for LLMs.

We validate our approach through three experiments. \textbf{First}, our expert-labeled dataset consistently improves the out-of-distribution (OOD) performance of mainstream post-training methods, clearly surpassing existing datasets without expert annotations. \textbf{Second}, we show that the Psy-Interpreter framework with dual-system training substantially outperforms standard generation or training methods across multiple diverse task types. \textbf{Third}, we show that models trained under our framework not only generalize better to unseen psychological tasks, but also possess self-annotation capabilities that enable continual learning, thereby closing the gap between expert supervision and autonomous psychological reasoning.

Together, these contributions form a unified and practical framework for advancing social-cognitive reasoning in LLMs through real-world data, psychologically informed learning, and efficient model optimization. Based on our experiments, we argue that explicit knowledge injection and the imitation of expert psychological thought patterns collectively endow the model with expert-level interpretive capabilities.

\section{Related Works}
\label{s2}

\subsection{LLMs in Psychological Tasks}

LLMs are increasingly evaluated on psychological tasks involving theory of mind and moral reasoning. Benchmarks show that even top models such as GPT-4o underperform humans in belief reasoning, indicating reliance on superficial cues \cite{xiao2025towards,shapira-etal-2024-clever}. Datasets such as SimpleToM, CogBench, and SocialIQa extend evaluations to emotion and social inference \cite{gu2024simpletom,coda2024cogbench,sap2019socialiqa}. In moral reasoning, LLMs may align with or even surpass humans in perceived ethical competence \cite{liu2025prompt,huang2023chatgpt}, though often via pattern matching. Empathy studies find LLMs capable of emotionally appropriate responses, occasionally preferred over human ones \cite{Ayers2023Comparing,sorin2024large}, albeit with limited contextual nuance \cite{yang2024emollm}. Open models still lag behind proprietary ones \cite{li2024quantifying}, highlighting the need for theory-informed benchmarks and training methods \cite{xie2024psydt,qiu-etal-2024-smile}.
\begin{figure*}[!t]
    \centering
    \includegraphics[width=\linewidth]{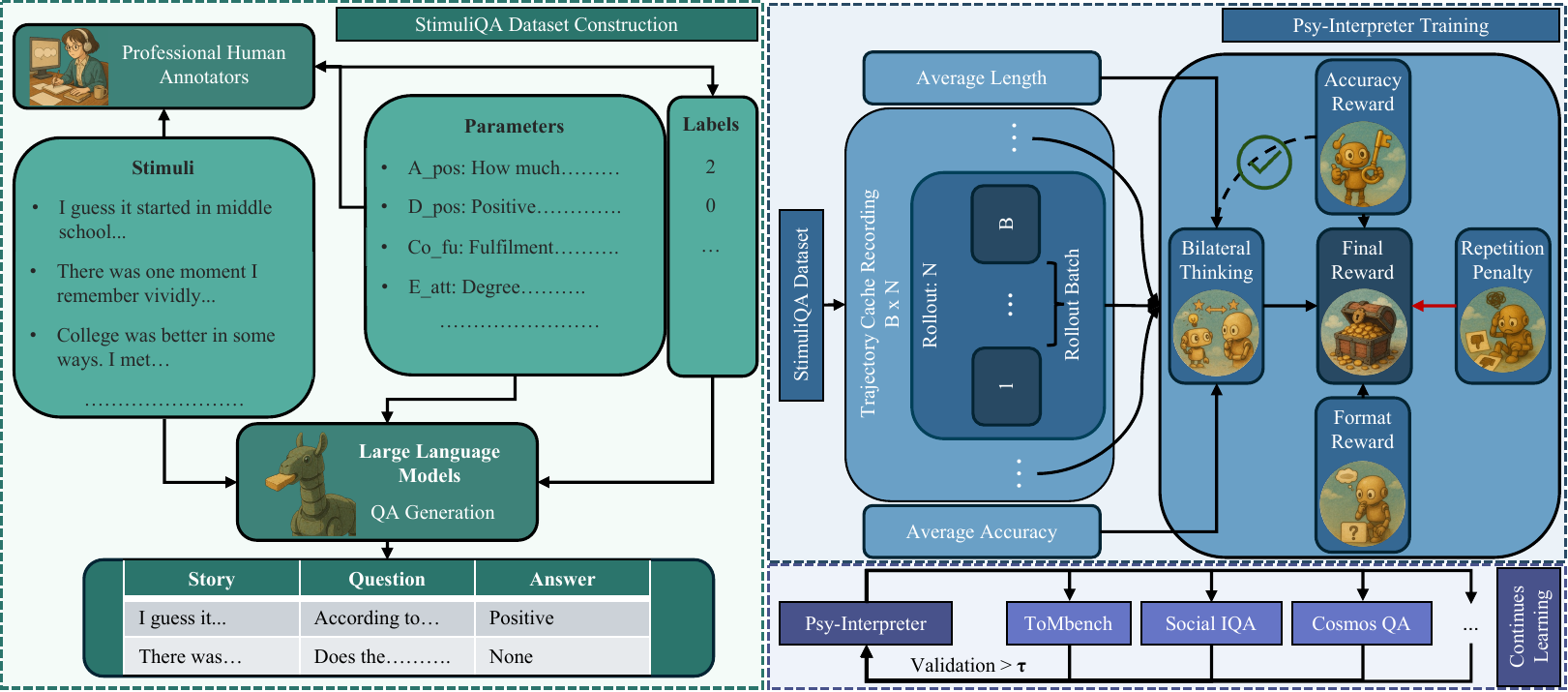}
    \caption{The framework comprises: \textit{StimuliQA}, stimuli with expert psychological labels; \textit{Psy-Interpreter}, a training framework tailored for psychological tasks; and \textit{Continual Learning}, demonstrating continual learning capability through self-labeling.}
    \label{fig:paper_structure}
\end{figure*}
\subsection{CoT and Reinforcement Learning}
Chain-of-Thought (CoT) prompting improves deliberate reasoning by encouraging stepwise thinking \cite{luo2025graphr1,shen2025flow}, and is often combined with RL to align outputs with human preferences \cite{Ouyang2022Training}. Process supervision and gradient-level feedback, such as in DeepSeek-R1, further enhance intermediate reasoning quality \cite{luo2025kbqao1}. Frameworks like Reflexion and Tree of Thoughts introduce self-evaluation and planning to strengthen coherence \cite{luo2025hypergraphrag,feng2025group}. Despite progress, challenges remain in balancing fluency with rigor and designing scalable feedback for ambiguous reasoning. CoT and RL provide a strong basis for aligning LLMs in psychological reasoning tasks and broader real-world applications.

\section{Methodology}
\label{s3}
This section outlines our methodology for enhancing psychological reasoning as shown in the Figure~\ref{fig:paper_structure}. We first construct \textbf{StimuliQA}, a dataset pairing human-labeled narratives with LLM-generated QA. We then propose \textbf{Bilateral Reinforcement Learning}, integrating token accuracy, format compliance, reasoning depth, and repetition control into a unified reward. A \emph{Trajectory Cache} stabilizes training via recent rollout tracking and bonus conditions. Finally, \textbf{Continuous Learning} enables refinement through confident predictions.

\subsection{Data Construction}

We built \textbf{StimuliQA} to support psychological reasoning by combining expert annotations, aiming to inject expert knowledge and guide LLMs toward human-like reasoning. Comparison with other datasets can be found in the Appendix.

\paragraph{Stimuli Collection and Annotation}  
We collected 3,280 real-life stimuli with emotional and social content. Human annotators labeled them based on variables above, and utilized LLMs to generate QA pairs from the labels, ensuring expert grounding and generative diversity.

\paragraph{Psychological Variable Design}  
The dataset comprises 58 variables across three dimensions: \textbf{Emotional Reactions} (29 variables), which represent affective and social-emotional responses grounded in Lazarus’s appraisal theory \cite{lazarus1991progress} , and further explored in settings as \cite{han2025narrative}; \textbf{Narrative Transformation} (12 variables), reflecting tone shifts and redemptive arcs, inspired by McAdams et al. \cite{mcadams2001bad}; and \textbf{Collective Psychology} (17 variables), covering indicators of self-worth and community connection based on Ryff \& Keyes’s model of psychological well-being \cite{ryff1995structure}. This variable structure enables exploration of how individuals process and narrate life experiences.

\subsection{Bilateral Reinforcement Learning}
We propose a Bilateral RL frame with a \emph{Trajectory Cache} that adopt \emph{Trajectory-aware GRPO} based on Group Relative Policy Optimization (GRPO) \cite{shao2024deepseekmath} and \emph{Bilateral Reward} stabilizes rewards and promotes structured reasoning.

\paragraph{Trajectory Cache}
\label{Trajectory}
Fixed rewards are often suboptimal across model scales or tasks. We use a \emph{Trajectory Cache} to track recent performance and adjust rewards by trend, stabilizing estimation. With $B$ batches and cache size $C$, the batch caches $B_c = B \times C$ summarize training dynamics.

\paragraph{T-GRPO Objective}
\label{Trajectory-aware Group Relative Policy Optimization}
The objective of T-GRPO (Trajectory-aware Group Relative Policy Optimization ) is defined as:
\begin{equation}
\begin{aligned}
&\!\mathcal{J}_{\text{T-GRPO}}(\theta)\! =\!
\frac{1}{B_cG} \sum_{b=1}^{B_c} \sum_{i=1}^{G} \frac{1}{|o_{b,i}|} \!\sum_{t=1}^{|o_{b,i}|}\!\bigg[
\min\bigg(r_{b,i,t}(\theta) \hat{A}_{b,i,t}, \\
&\qquad\text{clip}(r_{b,i,t}(\theta), 1 - \varepsilon, 1 + \varepsilon) \hat{A}_{b,i,t} \bigg)
\bigg]\!-\!\beta D_{\text{KL}}(\pi_\theta \,\|\, \pi_{\text{ref}}),
\end{aligned}
\end{equation}
where the importance ratio and normalized advantage are:
\begin{equation}
r_{b,i,t}(\theta) = \frac{\pi_\theta(o_{b,i,t} \mid q_b, o_{b,i,<t})}{\pi_{\theta_{\text{old}}}(o_{b,i,t} \mid q_b, o_{b,i,<t})}, \ 
\hat{A}_{b,i,t} = \frac{r^{\text{Final}}_{b,i} - \bar{r}}{\sigma_r + \epsilon},
\end{equation}
with $\bar{r}$ and $\sigma_r$ being the batch-level mean and standard deviation computed over all group-level final rewards $r^{\text{Final}}_{b,i}$ as defined in the \emph{Bilateral Reward} section.
\paragraph{Bilateral Reward}
\label{Bilateral Reward}
For our overall bilateral reward, each group $(b,i)$ is assigned a final reward for $r^{\text{Final}}_{b,i}$ listed here:
\begin{equation}
r^{\text{Final}}_{b,i} = w^{\text{F1}} \cdot r^{\text{F1}}_{b,i} + w^{\text{fmt}} \cdot r^{\text{fmt}}_{b,i} + r^{\text{BR}}_{b,i} -  r^{\text{rep}}_{b,i},
\end{equation}
where $w^{(\cdot)}$ denotes the weight and $r^{(\cdot)}_{b,i}$ denotes the reward term, which will be introduced in further detail below.

\paragraph{Answer Quality ($r^{\text{F1}}$)}
We compute $r^{\text{F1}}_{b,i}$ as the token-level F1 score between the predicted answer $\text{Ans}^{\text{pred}}_{b,i}$ the corresponding ground-truth answer $\text{Ans}^{\text{gold}}_{b,i}$. Formally: 
\begin{equation}
r^{\text{F1}}_{b,i} = \frac{2  P_{b,i}  R_{b,i}}{P_{b,i} + R_{b,i}},
\end{equation}
\begin{equation}
\ P_{b,i} = \frac{|y^{\text{pred}}_{b,i} \cap y^{\text{gold}}_{b,i}|}{|y^{\text{pred}}_{b,i}|},  \ 
R_{b,i} = \frac{|y^{\text{pred}}_{b,i} \cap y^{\text{gold}}_{b,i}|}{|y^{\text{gold}}_{b,i}|}, 
\end{equation}
where $y^{\text{pred}}_{b,i}$ \& $y^{\text{gold}}_{b,i}$ are token sets after normalization, including lowercase conversion, punctuation and space cleanup.

\paragraph{Format Compliance ($r^{\text{fmt}}$)} 
\label{fmt}
To ensure structural consistency, we define a binary reward for format correctness:
\begin{equation}
r^{\text{fmt}}_{b,i} = 
\begin{cases}
1, & \text{if } o_{b,i} \text{ matches predefined format constraints,} \\
0, & \text{otherwise.}
\end{cases}
\end{equation}
Format validation checks if the response is in the correct format (
\textnormal{\textless think\textgreater}...
\textnormal{\textless/think\textgreater}
\textnormal{\textless answer\textgreater}...
\textnormal{\textless/answer\textgreater}
).

\paragraph{Bilateral Reasoning ($r^{\text{BR}}$)}
To encourage informative yet concise reasoning, we introduce a \emph{bilateral reward} term that jointly considers relative reasoning length and answer quality. 
Let $L_{b,i} = |o_{b,i}|$ denote the output length for group $(b,i)$, and define the batch-level averages as:
\begin{equation}
\bar{L} = \frac{1}{B_cG} \sum_{b=1}^{B_c} \sum_{i=1}^{G} L_{b,i}, \quad
\bar{r}^{\text{F1}} = \frac{1}{B_cG} \sum_{b=1}^{B_c} \sum_{i=1}^{G} r^{\text{F1}}_{b,i}.
\end{equation}
We then compute the bilateral reward as:
\begin{equation}
r^{\text{BR}}_{b,i} =
\begin{cases}
\delta, & \frac{L_{b,i}}{\bar{L}} < \tau_{-} \ \text{and}\ r^{\text{F1}}_{b,i} > \bar{r}^{\text{F1}},\\[2pt]
\delta, & \frac{L_{b,i}}{\bar{L}} > \tau_{+} \ \text{and}\ r^{\text{F1}}_{b,i} > \bar{r}^{\text{F1}},\\[2pt]
0, & \text{otherwise,}
\end{cases}
\end{equation}
where $\tau_{-}$ and $\tau_{+}$ denote the lower and upper thresholds, respectively. This reward encourages the model to adopt an appropriate reasoning length, using concise responses for simpler cases and extended reasoning for complex ones. 

Further, comparing $r^{\text{F1}}_{b,i}$ with $\bar{r}^{\text{F1}}$ ensures that length-based rewards are granted only to outputs that are both accurate and reliably reasoned, preventing the model from receiving incentives for merely producing longer or shorter responses without demonstrating genuine understanding. 

\paragraph{Repetition Penalty ($r^{\text{rep}}$)}  
To discourage verbosity and repetition, we compute a penalty over the \textnormal{\textless think\textgreater} segment based on the ratio of repeated to total 4-grams:
\begin{equation}
r^{\text{rep}}_{b,i} = - \min\left(\tau_{\text{rep}}, \frac{\text{Repeat}_4(o^{\text{think}}_{b,i})}{\text{Total}_4(o^{\text{think}}_{b,i}) + \epsilon} \right),
\end{equation}
where $\tau_{\text{rep}}$ denotes the maximum penalty score for penalizes semantic redundancy and promotes meaningful reasoning.

\paragraph{Other Rewards} Additional general reward formulations-including the base reward, short reward, length reward and the length-based reward with repetition penalty (RP) were also utilized and systematically compared in the later section (see Section~\ref{Experiments}). Full details are provided in the Appendix. 

\subsection{Continuous Learning}

To enable \textbf{self-evaluation} and \textbf{self-improvement} under low-resource settings, we propose a simple framework for \textit{continuous learning}. The model not only generates predictions but also assesses their quality and improves over time, producing more \textbf{user-aligned} and \textbf{psychologically appropriate} responses. We define the learning criterion as:
\begin{equation}
\text{self\_train}(x) \iff \text{valid}(x) \land \text{confidence}(x) > \tau,
\end{equation}
where $x$ is a model output, $\text{valid}(x)$ verifies format and $\text{confidence}(x)$ is the model’s estimated confidence score.
\section{Experiments}
\label{Experiments}
This section is organized around five core research questions (RQs):
\textbf{RQ1:} Can human-labeled psychological data improve model generalization under various training methods?
\textbf{RQ2:} Do tailored reward functions enhance recognition of explicit and latent psychological states?
\textbf{RQ3:} Does Psy-Interpreter perform well on OOD datasets?
\textbf{RQ4:} Is continual learning generally effective?
\textbf{RQ5:} Can Psy-Interpreter reliably identify if a psychological question requires reasoning?

\subsection{Experimental Setup}
\begin{figure}[!t]
    \centering
    \includegraphics[width=\linewidth]{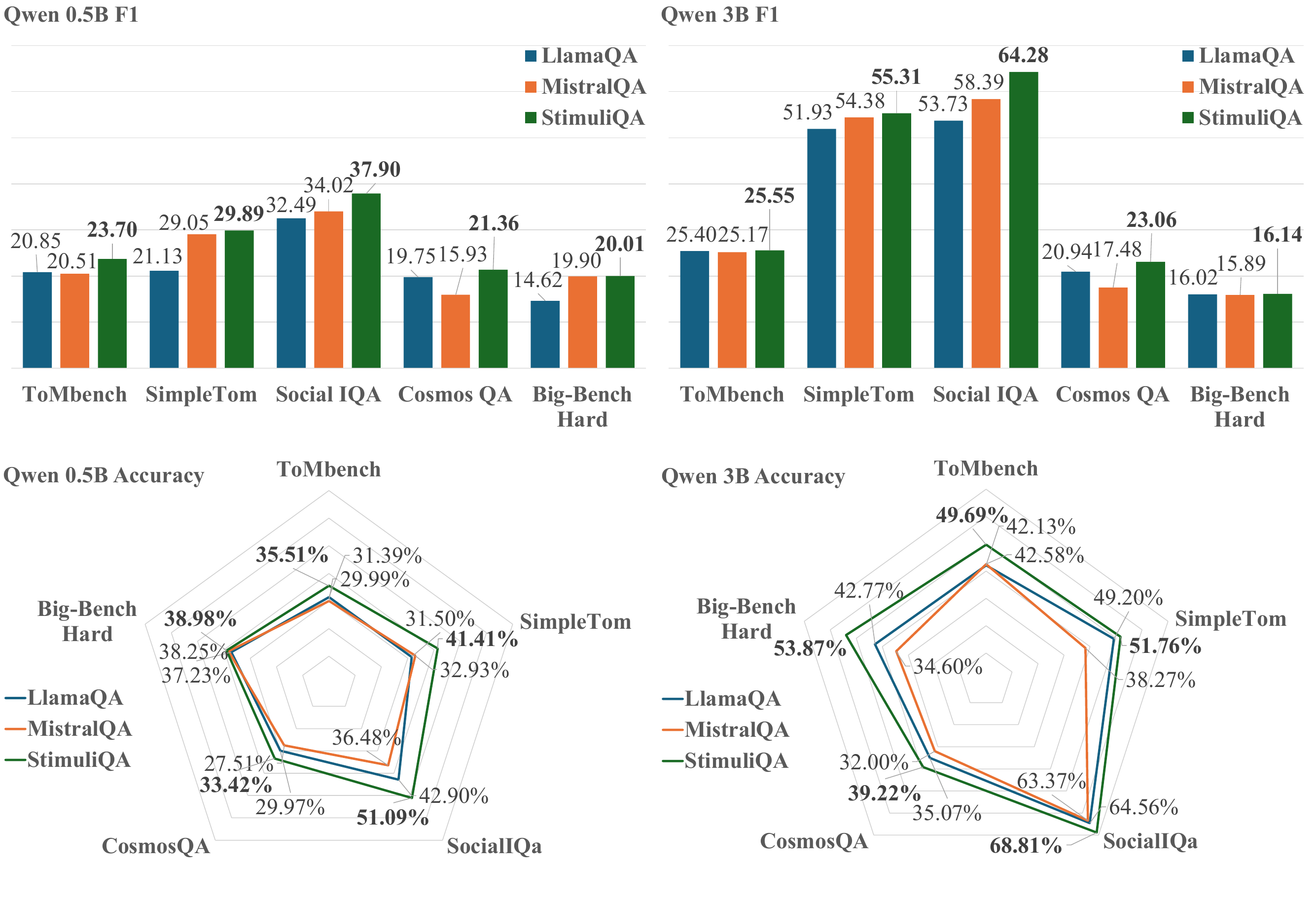}
    \caption{Comparison of GRPO training on StimuliQA and two other training datasets on Qwen0.5B and 3B. Full table with Qwen1.5B’s results is located in the Appendix.}
    \label{fig:grpo}
\end{figure}

\paragraph{Datasets} We used six different datasets: our StimuliQA and five OOD sets—ToMbench~\cite{chen2024tombench}, SimpleToM~\cite{gu2024simpletom}, SocialIQa~\cite{sap-etal-2019-social}, CosmosQA~\cite{huang-etal-2019-cosmos}, and selected BIG-Bench Hard~\cite{suzgun-etal-2023-challenging} subsets (disambiguation\_qa, formal\_fallacies, causal\_judgement). For each, we extracted the correct option to construct QA pairs grounded in narrative context. As SimpleToM and SocialIQa answers were difficult to connect to their questions, we appended them to the question text. We sampled 200 training and 100 test examples per psychological parameter, excluding four low-frequency ones (C\_acol, C\_agov, C\_anet, C\_apol), yielding 10,800 training and 5,400 test instances left for training and testing proposes.
\paragraph{Metrics} 
Consistent with prior QA research like\cite{deutsch2021towards} and \cite{su2019generalizing}, 
we use F1 score to evaluate our dataset. For OOD benchmarks, we adopt F1-based accuracy: each output is matched to the option with highest F1. If multiple options tie, we return E (“Not Applicable”); accuracy is computed on these selections.
\paragraph{Implementation Details}
We conduct three experiments. First, we train models on our dataset using two training paradigms (Table~\ref{tab:sft}, Figure~\ref{fig:grpo}) and compare their performance against two constructed comparison datasets, LlamaQA and MistralQA. Second, we train Qwen models on StimuliQA with varied reward functions and decoding strategies, evaluated on the held-out test set (Table~\ref{tab:rewards}). Third, we label logical reasoning samples on StimuliQA and use them to SFT-train Qwen models, thereby injecting knowledge without changing the generation format~\cite{mecklenburg2024injecting}. As Reinforcement Learning from Human Feedback (RLHF) primarily re-ranks existing knowledge~\cite{christiano2017deep, yue2025does, li2025generalist}, we apply our RL framework to the SFT model to produce \textbf{Psy-Interpreter}, evaluated on five OOD datasets and further tested with a continual learning module. Full details are provided in the Appendix.

\begin{table*}[!t]
\centering
\small
\begin{tabular}{ccccccccccc}
\toprule
\multirow{2}{*}{\textbf{Training Sets}} &
  \multicolumn{2}{c}{\textbf{ToMbench}} &
  \multicolumn{2}{c}{\textbf{SimpleTom}} &
  \multicolumn{2}{c}{\textbf{SocialIQa}} &
  \multicolumn{2}{c}{\textbf{CosmosQA}} &
  \multicolumn{2}{c}{\textbf{BIG-Bench Hard}} \\
\cmidrule(lr){2-3} \cmidrule(lr){4-5} \cmidrule(lr){6-7} \cmidrule(lr){8-9} \cmidrule(lr){10-11}
& F1 & Acc & F1 & Acc & F1 & Acc & F1 & Acc & F1 & Acc \\
\midrule
\multicolumn{11}{c}{\textbf{Qwen2.5-0.5B-Instruct}} \\
\midrule
LlamaQA      & 11.95 & 26.89\% & 15.37 & 32.03\% & 15.55 & 30.60\% & 12.29 & 28.21\% & 11.06 & 40.47\% \\
MistralQA & 10.01 & 21.33\% & 20.43 & 26.88\% & 22.79 & 32.70\% & 10.30 & 24.29\% & 14.45 & 39.74\% \\
StimuliQA     & \textbf{16.61} & \textbf{34.13\%} & \textbf{22.32} & \textbf{53.59\%} & \textbf{25.37} & \textbf{40.74\%} & \textbf{12.82} & \textbf{30.08\%} & \textbf{16.23} & \textbf{48.91\%} \\
\midrule
\multicolumn{11}{c}{\textbf{Qwen2.5-1.5B-Instruct}} \\
\midrule
LlamaQA      & 12.14 & 27.52\% & 20.03 & 34.93\% & 21.80 & 40.28\% & 12.43 & 28.74\% & 12.39 & 44.10\% \\
MistralQA & 10.27 & 22.62\% & 25.02 & 33.74\% & 27.31 & 37.87\% & 11.31 & 26.33\% & 14.50 & 39.88\% \\
StimuliQA     & \textbf{18.26} & \textbf{37.94\%} & \textbf{36.21} & \textbf{49.84\%} & \textbf{27.32} & \textbf{41.91\%} & \textbf{14.23} & \textbf{30.92\%} & \textbf{23.51} & \textbf{56.04\%} \\
\midrule
\multicolumn{11}{c}{\textbf{Qwen2.5-3B-Instruct}} \\
\midrule
LlamaQA      & 15.91 & 32.69\% & 18.48 & 26.16\% & 34.62 & 49.59\% & 15.11 & 31.96\% & 13.63 & 44.98\% \\
MistralQA & 14.88 & 30.70\% & 35.02 & 33.33\% & 35.28 & 40.84\% & 12.69 & 27.34\% & 19.33 & 41.05\% \\
StimuliQA     & \textbf{22.06} & \textbf{42.38\%} & \textbf{37.62} & \textbf{56.44\%} & \textbf{43.40} & \textbf{53.74\%} & \textbf{15.99} & \textbf{33.17\%} & \textbf{19.74} & \textbf{54.88\%} \\
\bottomrule
\end{tabular}
\caption{Comparison of SFT training on StimuliQA and two other datasets by both accuracy and F1 score.}
\label{tab:sft}
\end{table*}

\begin{table*}[t]
\small
\centering
\begin{tabular}{lccccccc}
\toprule
\textbf{Model} &
  \textbf{\begin{tabular}[c]{@{}c@{}}Specific\\ Emotions\end{tabular}} &
  \textbf{\begin{tabular}[c]{@{}c@{}}Targeted\\ Emotions\end{tabular}} &
  \textbf{\begin{tabular}[c]{@{}c@{}}Direction of \\ Change\end{tabular}} &
  \textbf{\begin{tabular}[c]{@{}c@{}}Self\\ Identification\end{tabular}} &
  \textbf{\begin{tabular}[c]{@{}c@{}}Needs\\ Fulfilment\end{tabular}} &
  \textbf{\begin{tabular}[c]{@{}c@{}}National \\ Theme\end{tabular}} &
  \textbf{Overall} \\
\midrule
\multicolumn{8}{c}{\textbf{Qwen2.5-0.5B-Instruct}} \\
Naïve & 10.72 & 9.20 & 14.60 & 13.78 & 10.58 & 13.76 & 11.46 \\
CoT & 10.95(+0.23) & 10.43(+1.23) & 14.51(-0.09) & 15.30(+1.52) & 11.42(+0.84) & 14.11(+0.35) & 12.31(+0.85) \\
SFT & \textbf{14.08(+3.36)} & \textbf{19.70(+10.5)} & \textbf{29.27(+14.67)} & \textbf{21.29(+7.51)} & \textbf{22.07(+11.49)} & \textbf{29.79(+16.03)} & \textbf{21.70(+10.24)} \\
\midrule
Basic R1 & 25.25 & 27.11 & 27.91 & 32.78 & 32.59 & 30.62 & 29.23 \\
Length Reward & 25.49(+0.24) & 26.83(-0.28) & 28.40(+0.49) & 32.80(+0.02) & 31.86(-0.73) & 27.81(-2.81) & 28.63(-0.60) \\
Short Reward & 22.61(-2.64) & 28.90(+1.79) & 27.82(-0.09) & 34.52(+1.74) & 33.80(+1.21) & 28.50(-2.12) & 29.49(+0.26) \\
Length w RP & 26.70(+1.45) & 28.49(+1.38) & 29.85(+1.94) & 34.59(+1.81) & 33.67(+1.08) & 30.18(-0.44) & 30.37(+1.14) \\
\textbf{BR} & \textbf{29.02(+3.77)} & \textbf{31.20(+4.09)} & \textbf{38.70(+10.79)} & \textbf{37.84(+5.06)} & \textbf{39.97(+7.38)} & \textbf{38.74(+8.12)} & \textbf{34.97(+5.74)} \\
\midrule
\multicolumn{8}{c}{\textbf{Qwen2.5-1.5B-Instruct}} \\
Naïve & 11.88 & 10.68 & 18.06 & 15.64 & 13.13 & 16.67 & 13.46 \\
CoT & 14.87(+2.99) & 15.07(+4.39) & 22.59(+4.53) & 19.09(+3.45) & 17.68(+4.55) & 20.18(+3.51) & 17.37(+3.91) \\
SFT & \textbf{19.13(+7.25)} & \textbf{21.48(+10.8)} & \textbf{33.00(+14.94)} & \textbf{31.32(+15.68)} & \textbf{29.43(+16.30)} & \textbf{31.58(+14.91)} & \textbf{26.27(+12.81)} \\
\midrule
Basic R1 & 34.29 & 30.57 & 37.03 & 35.28 & 38.22 & 36.81 & 34.46 \\
Length Reward & 33.26(-1.03) & 30.91(+0.34) & 35.94(-1.09) & 35.89(+0.61) & 38.00(-0.22) & 36.75(-0.06) & 34.41(-0.05) \\
Short Reward & 34.81(+0.52) & 30.56(-0.01) & 36.92(-0.11) & 37.00(+1.72) & 38.78(+0.56) & 38.23(+1.42) & 35.14(+0.68) \\
Length w RP & 34.90(+0.61) & 32.38(+1.81) & 37.05(+0.02) & 38.01(+2.73) & 40.54(+2.32) & 39.25(+2.44) & 36.32(+1.86) \\
\textbf{BR} & \textbf{38.45(+4.16)} & \textbf{36.00(+5.43)} & \textbf{41.95(+4.92)} & \textbf{42.09(+6.81)} & \textbf{43.88(+5.66)} & \textbf{42.52(+5.71)} & \textbf{39.98(+5.52)} \\
\midrule
\multicolumn{8}{c}{\textbf{Qwen2.5-3B-Instruct}} \\
Naïve & 13.44 & 13.27 & 19.97 & 16.20 & 15.19 & 18.68 & 15.34 \\
CoT & 13.86(+0.42) & 15.11(+1.84) & 21.60(+1.63) & 17.65(+1.45) & 16.27(+1.08) & 20.03(+1.35) & 16.70(+1.36) \\
SFT & \textbf{22.45(+9.01)} & \textbf{22.04(+8.77)} & \textbf{34.47(+14.50)} & \textbf{31.68(+15.48)} & \textbf{27.24(+10.97)} & \textbf{32.73(+14.05)} & \textbf{26.95(+11.61)} \\
\midrule
Basic R1 & 34.17 & 31.23 & 36.81 & 36.04 & 37.98 & 38.71 & 35.04 \\
Length Reward & 33.02(-1.15) & 31.56(+0.33) & 37.90(+1.09) & 36.77(+0.73) & 38.78(+0.80) & 37.86(-0.85) & 35.14(+0.10) \\
Short Reward & 35.46(+1.29) & 31.28(+0.05) & 38.02(+1.21) & 37.84(+1.80) & 40.38(+2.40) & 37.72(-0.99) & 35.81(+0.77) \\
Length w RP & 35.50(+1.33) & 33.56(+2.33) & 39.88(+3.07) & 39.02(+2.98) & 41.79(+3.81) & 39.80(+1.09) & 37.39(+2.35) \\
\textbf{BR} & \textbf{39.17(+5.00)} & \textbf{36.25(+5.02)} & \textbf{41.87(+5.06)} & \textbf{42.03(+5.99)} & \textbf{44.44(+6.46)} & \textbf{43.38(+4.67)} & \textbf{40.38(+5.34)} \\
\bottomrule
\end{tabular}
\caption{Comparison of different training and generation methods on our StimuliQA dataset across six psychological dimensions from our StimuliQA. Our BR reward consistently achieves the best overall performance across all six domains.}
\label{tab:rewards}
\end{table*}

\begin{table*}[t]
\centering
\small
\begin{tabular}{llcccccccccc}
\toprule
\multicolumn{2}{c}{\textbf{Models}} &
  \multicolumn{2}{c}{\textbf{ToMbench}} &
  \multicolumn{2}{c}{\textbf{SimpleTom}} &
  \multicolumn{2}{c}{\textbf{SocialIQa}} &
  \multicolumn{2}{c}{\textbf{CosmosQA}} &
  \multicolumn{2}{c}{\textbf{BIG-Bench Hard}} \\
\cmidrule(lr){3-4} \cmidrule(lr){5-6} \cmidrule(lr){7-8} \cmidrule(lr){9-10} \cmidrule(lr){11-12}
\multicolumn{2}{c}{} &
  F1 &
  Acc &
  F1 &
  Acc &
  F1 &
  Acc &
  F1 &
  Acc &
  F1 &
  Acc \\
\midrule
\multicolumn{2}{c}{\textbf{GPT-4 nano}} &
  22.11 & 46.22\% & 67.03 & 54.34\% & 57.03 & 68.03\% & \textbf{15.82} & 35.00\% & 9.30 & 41.75\% \\
\multicolumn{2}{c}{\textbf{Claude 3 Haiku}} &
  18.81 & 49.65\% & 19.32 & 37.31\% & 15.94 & 52.64\% & 11.28 & 33.27\% & 7.07 & 40.15\% \\
\multicolumn{2}{c}{\textbf{DeepSeek Reasoner}} &
  26.63 & \textbf{56.02\%} & \textbf{78.98} & \textbf{69.17\%} & \textbf{63.93} & \textbf{70.36\%} & 13.13 & 35.23\% & \textbf{40.52} & \textbf{69.93\%} \\
\multicolumn{2}{c}{\textbf{Llama 3.3 70B Instruct}} &
  \textbf{27.42} & 52.24\% & 76.43 & 61.46\% & 60.91 & 69.07\% & 15.27 & \textbf{37.19\%} & 17.61 & 69.34\% \\
\multicolumn{2}{c}{\textbf{Mistral 8$\times$7B Instruct}} &
  12.03 & 34.46\% & 12.06 & 42.52\% & 12.14 & 42.23\% & 7.49 & 24.66\% & 5.39 & 26.42\% \\
\midrule
\multicolumn{12}{c}{\textbf{Qwen2.5-0.5B-Instruct}} \\
\multicolumn{2}{c}{Simple RL Zoo} &
  13.38 & 14.42\% & 12.89 & 6.63\% & 16.56 & 12.23\% & 10.33 & 5.22\% & 9.12 & 6.86\% \\
\multicolumn{2}{c}{NuExtract} &
  12.69 & 7.21\% & 12.20 & 1.39\% & 14.44 & 2.54\% & 8.02 & 2.50\% & 6.83 & 2.04\% \\
\multicolumn{2}{c}{New Merges Serialization} &
  15.21 & 23.27\% & 17.30 & 13.86\% & 15.14 & 22.69\% & 9.80 & 22.69\% & 8.04 & 6.13\% \\
\multicolumn{2}{c}{Psy-Interpreter} &
  25.27 & 40.66\% & 37.63 & 65.10\% & 43.01 & 51.30\% & 23.58 & 38.69\% & 23.05 & 48.76\% \\
\multicolumn{2}{c}{Psy-Interpreter-SFT} &
  \textbf{62.28} & \textbf{58.82\%} & \textbf{85.25} & \textbf{65.45\%} & \textbf{76.58} & \textbf{74.04\%} & \textbf{34.43} & \textbf{43.45\%} & \textbf{62.41} & \textbf{56.06\%} \\
\midrule
\multicolumn{12}{c}{\textbf{Qwen2.5-1.5B-Instruct}} \\
\multicolumn{2}{c}{Simple RL Zoo} &
  3.36 & 4.51\% & 9.02 & 6.51\% & 4.33 & 7.31\% & 2.91 & 2.50\% & 3.98 & 8.03\% \\
\multicolumn{2}{c}{Nemotron Reasoning} &
  2.02 & 4.65\% & 6.38 & 6.10\% & 7.30 & 10.21\% & 1.69 & 5.26\% & 6.98 & 9.78\% \\
\multicolumn{2}{c}{DeepSeek-R1 Distilled} &
  11.39 & 22.64\% & 26.80 & 26.68\% & 9.25 & 22.49\% & 7.33 & 12.14\% & 2.73 & 13.28\% \\
\multicolumn{2}{c}{Psy-Interpreter} &
  26.54 & 49.30\% & 39.02 & 59.31\% & 56.40 & 63.94\% & 23.61 & 41.34\% & 22.33 & 50.07\% \\
\multicolumn{2}{c}{Psy-Interpreter-SFT} &
  \textbf{65.94} & \textbf{62.25\%} & \textbf{84.95} & \textbf{66.26\%} & \textbf{80.55} & \textbf{78.86\%} & \textbf{36.37} & \textbf{45.68\%} & \textbf{64.04} & \textbf{58.69\%} \\
\midrule
\multicolumn{12}{c}{\textbf{Qwen2.5-3B-Instruct}} \\
\multicolumn{2}{c}{Transformer RL} &
  16.25 & 18.86\% & 15.55 & 9.56\% & 25.20 & 39.33\% & 11.14 & 6.53\% & 9.33 & 5.26\% \\
\multicolumn{2}{c}{Tulu Trained} &
  21.54 & 22.67\% & 33.83 & 16.97\% & 46.24 & 39.12\% & 14.44 & 24.28\% & 34.92 & 29.34\% \\
\multicolumn{2}{c}{Raspberry} &
  15.46 & 15.96\% & 33.92 & 24.35\% & 28.93 & 26.06\% & 10.36 & 3.65\% & 13.23 & 6.57\% \\
\multicolumn{2}{c}{Psy-Interpreter} &
  28.17 & 51.33\% & 56.83 & 58.73\% & 65.54 & 71.09\% & 25.60 & 44.33\% & 18.87 & 56.06\% \\
\multicolumn{2}{c}{Psy-Interpreter-SFT} &
  \textbf{66.33} & \textbf{58.36\%} & \textbf{84.98} & \textbf{67.92\%} & \textbf{82.82} & \textbf{81.24\%} & \textbf{36.88} & \textbf{47.56\%} & \textbf{65.86} & \textbf{60.88\%} \\
\bottomrule
\end{tabular}
\caption{Performance of our Psy-Interpreter and continual learning with SFT on all the OOD datasets, compared against strong commercial models and open-source baselines, with consistent gains across all five datasets.}
\label{tab:psy-eval}
\end{table*}

\subsection{RQ1: Effect of Human-Labeled Data}

One of the central hypotheses in our work is that human-labeled data, particularly when annotated by trained psychology students, can inject high-quality domain knowledge into compact models during SFT. We evaluate this hypothesis by comparing models trained on our \textbf{StimuliQA} dataset against those trained on synthetic datasets generated by two powerful large models: Llama 3.3-70B and Mistral 8$\times$7B. All datasets were used to train Qwen2.5-Instruct models of three different sizes (0.5B, 1.5B, 3B) via SFT, and evaluated on five diverse OOD benchmarks: \textbf{ToMbench}, \textbf{SimpleToM}, \textbf{SocialIQa}, \textbf{CosmosQA}, and \textbf{BIG-Bench Hard}. Where we have the final results displayed in Table~\ref{tab:sft} and Figure~\ref{fig:grpo}. This comparison reveals performance gaps and tests the transferability of psychological reasoning across model sizes and tasks. These benchmarks span theory of mind, social commonsense, and logical reasoning, allowing for a comprehensive assessment of domain generalization and model capabilities.

As shown in Table~\ref{tab:sft}, models trained on \textbf{StimuliQA} consistently outperform those trained on synthetic data across all model sizes and benchmarks, with especially strong gains in the 3B setting. For instance, on \textbf{SimpleToM}, Qwen2.5-3B achieves 37.62 F1 and 56.44\% accuracy, far exceeding Llama 3.3 (18.48/26.16\%) and Mistral 8$\times$7B (35.02/33.33\%). These substantial margins further reinforce our claim that: human-labeled data conveys more nuanced and structured psychological knowledge than synthetic outputs. The consistent OOD improvements indicate this knowledge is indeed generalizable, highlighting the professional quality and domain relevance of our dataset. Annotations by trained psychology students provide valuable domain expertise essential for effective model training. When used in GRPO training (Figure~\ref{fig:grpo}), human-labeled data still offers an advantage.

\subsection{RQ2: Impact of Reward Design}

\begin{figure}[!t]
    \centering
    \includegraphics[width=\linewidth]{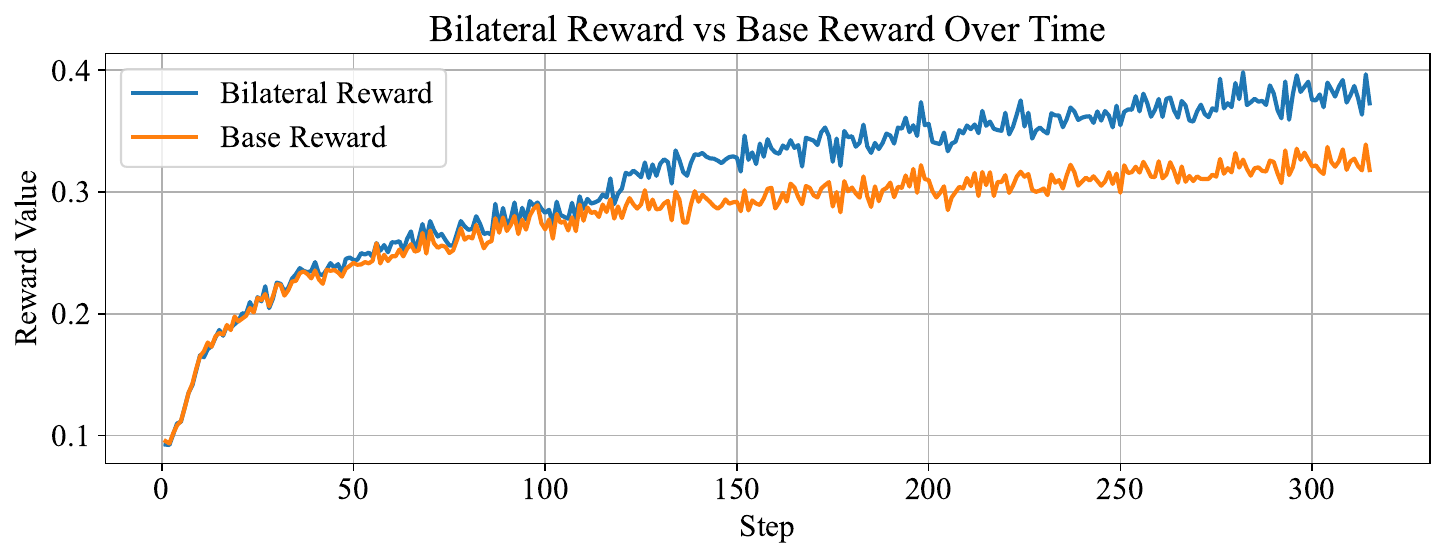}
    \caption{Base and Bilateral reward Training comparison.}
    \label{fig:train}
\end{figure}

Our results in Table~\ref{tab:rewards} show that psychologically grounded reward functions substantially enhance model performance, and Figure~\ref{fig:train} also shows a significant training accuracy boost. In particular, the BR reward, which integrates structure, emotion sensitivity, and length normalization, consistently outperforms baselines like Basic R1 or single-aspect rewards. For instance, BR raises overall F1 on Qwen2.5-1.5B from 34.46 to 39.98, with +5.66 and +5.71 F1 gains in \textit{Needs Fulfilment} and \textit{National Theme}, respectively. These gains suggest BR promotes not only token-level accuracy but also deeper emotional and moral reasoning. Beyond absolute improvements, BR yields more balanced results across all psychological dimensions, unlike length-based or repetition-penalized rewards, which often improve certain aspects at the cost of degrading others. This highlights BR’s ability to align model behavior with complex, multi-faceted psychological goals.

Moreover, BR scales well across model sizes, offering +5.34 F1 on Qwen-3B and +5.74 on Qwen-0.5B, making it suitable for compute-constrained settings. Finally, BR-trained models generate more coherent moral reasoning, richer self-reflection, and clearer emotional distinctions indicating a shift toward more theory-aligned internal representations.

\subsection{RQ3: Generalization of Psy-Interpreter}

As shown in Table~\ref{tab:psy-eval}, \textbf{Psy-Interpreter} demonstrates strong zero-shot generalization across five diverse OOD benchmarks after injecting knowledge via SFT and GRPO training with our \textbf{StimuliQA} dataset. Despite being trained without any direct supervision on these test sets, it consistently outperforms other baselines such as Simple RL Zoo, NuExtract, and DeepSeek-R1 Distilled across all model sizes. Notably, the 3B variant achieves 28.17 F1 on \textbf{ToMbench}, 56.83 on \textbf{SimpleToM}, and 65.54 on \textbf{SocialIQa}, marking substantial gains over both in-house and open-source compact models. These results suggest that Psy-Interpreter successfully internalizes psychologically grounded knowledge during training, enabling it to reason beyond its original domain.

\subsection{RQ4: Effectiveness of Continual Learning}
Table~\ref{tab:psy-eval} shows that our continual learning framework (\textbf{Psy-Interpreter-SFT}) achieves substantial and consistent improvements across all five OOD psychological benchmarks. For the 0.5B model, accuracy rises from 40.66\% to \textbf{58.82\%} on ToMbench, 51.30\% to \textbf{74.04\%} on SocialIQa, and 38.69\% to \textbf{43.45\%} on CosmosQA. F1 scores increase even more sharply, e.g., 25.27 to \textbf{62.28} on ToMbench and 23.05 to \textbf{62.41} on BIG-Bench Hard, indicating better correctness and closer alignment with human-preferred reasoning. The 1.5B and 3B variants maintain these gains, reaching up to 82.82 F1 on SocialIQa and consistently surpassing their non-SFT counterparts. Despite their compact size, our models rival or exceed larger commercial LLMs: \textbf{Psy-Interpreter-SFT (3B)} outperforms GPT-4 nano (57.03) and Claude 3 Haiku (15.94), and matches or surpasses DeepSeek Reasoner on most tasks.

Despite their smaller scale, our models often remarkably match or notably surpass larger commercial LLMs. \textbf{Psy-Interpreter-SFT (3B)} achieves an F1 of 82.82 on SocialIQa, exceeding GPT-4 nano (57.03) and Claude 3 Haiku (15.94), and outperforming DeepSeek Reasoner on most tasks.

\subsection{RQ5: Length Distribution Comparison}
\begin{figure}[!t]
    \centering
    \includegraphics[width=\linewidth]{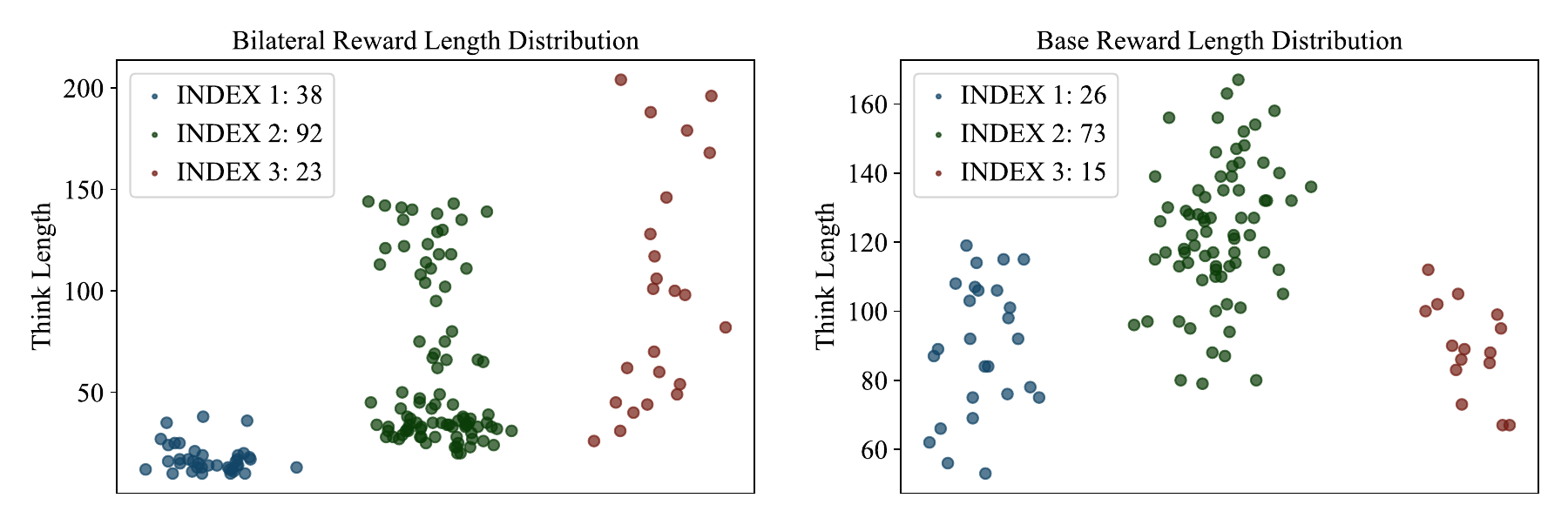}
    \caption{The figure shows the distribution of reasoning lengths under the bilateral reward and the base reward.}
    \label{fig:reason_length}
\end{figure}

Figure~\ref{fig:reason_length} illustrates the reasoning length distributions under the bilateral and base reward settings on the \emph{Unexpected Outcome Test} from ToMbench, which comprises 300 samples with an intuitive question-difficulty hierarchy. 
We only include samples with non-zero F1 scores, as these responses provide reasoning chains that are more representative and interpretable than those from entirely incorrect outputs.
Index~1 corresponds to simple stories with simple questions, Index~2 to the same stories with harder questions, and Index~3 to more complex stories with direct questions.
The figure also reports the total number of valid samples under each reward configuration.
As shown, the bilateral reward yields a more structured and interpretable distribution: 
Index~1 responses remain concise, while Index~2 and Index~3 exhibit longer and more differentiated reasoning chains. 
In contrast, the base reward produces a noisier distribution with reduced separation between simple and complex QA pairs, occasionally generating overly long reasoning even for direct emotion predictions. 
This indicates that our bilateral reward better aligns reasoning length with task difficulty, enhancing overall interpretability and efficiency. 
Moreover, the consistent accuracy gains in Tables~\ref{tab:rewards} and \ref{tab:psy-eval} underscore the importance of carefully controlling reasoning length in complex psychological reasoning. 
In addition, our reward method was able to generate more correct samples at each difficulty level, yielding 12 additional samples for simple emotion prediction and 19 and 8 more for the harder ones, respectively, validating its effectiveness.

Figure~\ref{fig:think_think} shows a reasoning samples with BR. More detailed comparisons and analysis can be found in the Appendix where we argued that our training framework successfully improved Emotional Quotient (EQ) of LLMs. 

\section{Conclusion}
We first introduce the expert-annotated dataset StimuliQA and Psy-Interpreter, an RL framework firmly grounded in psychological theory that enhances emotional and cognitive reasoning in LLMs. Leveraging high-quality human-labeled data with structured RL substantially improves social-cognitive reasoning in compact models, yielding consistent gains across five diverse OOD benchmarks. Our Bilateral Reward aligns reasoning length with task complexity and improves accuracy, while Psy-Interpreter-SFT with continual learning module effectively narrows the gap with larger commercial LLMs.

\bibliography{aaai2026}
\appendix
\section{Training Parameters}
For all our experiments, we used four NVIDIA RTX A6000 with memory usage of 49140MiB. And driver version of 550.144.03 and CUDA version of 12.4.
\subsection{Prompts for Testing}

\paragraph{CoT-related Prompt.} 
\texttt{[content | trim]} A conversation between User and Assistant.  
The user provides a “Stimuli:”, and the Assistant reads it and answers the “Question:”.  
The Assistant first thinks about the reasoning process internally, then provides the answer to the user.  
The reasoning and answer are enclosed in the following tags:

\begin{itemize}
    \item Reasoning: \textnormal{\textless think\textgreater} ... \textnormal{\textless/think\textgreater}
    \item Answer: \textnormal{\textless answer\textgreater} ... \textnormal{\textless/answer\textgreater}
\end{itemize}

Example:  
\textnormal{\textless think\textgreater} reasoning process here \textnormal{\textless/think\textgreater}  
\textnormal{\textless answer\textgreater} answer here \textnormal{\textless/answer\textgreater}

\paragraph{Direct Generation Prompt.} 
\texttt{[content | trim]} A conversation between User and Assistant.  
The user provides a “Stimuli:”, and the Assistant reads it and answers the “Question:”.  
The Assistant only provides the answer, enclosed as:

\begin{itemize}
    \item \textnormal{\textless answer\textgreater} answer here \textnormal{\textless/answer\textgreater}
\end{itemize}

\paragraph{Topic Generation Prompt.} 
Refer to the following topics as conceptual anchors.  
Generate as many psychologically or contextually related topics as possible while preserving thematic coherence:

\begin{verbatim}
["sample topics"]
\end{verbatim}

\paragraph{Dataset Generation Prompt.} 
You are an AI assistant that generates one question–answer pair based on a short Stimuli.  
The question should assess the human-labeled psychological parameter \texttt{[column\_name]} (\texttt{[description]}),  
with value range \texttt{[range]}. Ensure that the question:

\begin{itemize}
    \item Directly relates to the Stimuli,
    \item Evaluates the concept of \texttt{[category]},
    \item Follows the exact format:
\end{itemize}

\begin{verbatim}
Question: <Insert question here>
Answer: <Insert answer here>
Reason: <Explain why the answer is correct>
\end{verbatim}

Language should be clear, concise, and include plausible distractor options.

\paragraph{Comparison Dataset Generation Prompt.} 
Generate a Stimuli based on \texttt{([topic])}, then generate the Q\&A pair for  
psychological parameter \texttt{[column\_name]} (\texttt{[description]}) in the same format as above.

\paragraph{Reasoning CoT Generation Prompt.} 
Given a psychological prompt, generate a step-by-step reasoning process only.  

\begin{itemize}
    \item Do \textbf{not} output the final answer.
    \item Do \textbf{not} include \textless think\textgreater or \textless answer\textgreater tags.
    \item If it is a direct emotion prediction, use only a few words.
    \item Generate reasoning for \texttt{[QA]} with the expected answer \texttt{[Answer]}.
\end{itemize}

\subsection{Other rewards}
For our base reward, each group $(b,i)$ is assigned a final reward $r^{\text{Final}}_{b,i}$ as follows:
\begin{equation}
r^{\text{Final}}_{b,i} = w^{\text{F1}} \cdot r^{\text{F1}}_{b,i} + w^{\text{fmt}} \cdot r^{\text{fmt}}_{b,i}
\end{equation}

All components are aligned with the method section above. For the length or short reward, each group $(b,i)$ is assigned a final reward $r^{\text{Final}}_{b,i}$ as follows:
\begin{equation}
r^{\text{Final}}_{b,i} = w^{\text{F1}} \cdot r^{\text{F1}}_{b,i} + w^{\text{fmt}} \cdot r^{\text{fmt}}_{b,i} + w^{\text{length/short}} \cdot r^{\text{length/short}}_{b,i}
\end{equation}

For the length reward, we have the condition:
\begin{equation}
r^{\text{length}}_{b,i} \!=\!
\begin{cases}
\log\left(1 + \frac{r^{\text{F1}}_{b,i}}{\bar{r}^{\text{F1}}} \right)\!\!\cdot\!\!\frac{L_{b,i}}{\bar{L}}, \!\!\!\!\!\!\!& \frac{L_{b,i}}{\bar{L}}\!\le\! L_{+} \\
0, & \text{otherwise}
\end{cases}
\end{equation}

For the short reward, we have the condition:
\begin{equation}
r^{\text{short}}_{b,i} \!=\!
\begin{cases}
\log\left(1 + \frac{r^{\text{F1}}_{b,i}}{\bar{r}^{\text{F1}}} \right)\!\!\cdot\!\!\frac{L_{b,i}}{\bar{L}}, \!\!\!\!\!\!\!& \frac{L_{b,i}}{\bar{L}}\!\ge\! S_{-} \\
0, & \text{otherwise}
\end{cases}
\end{equation}

For the length reward with repetition penalty, each group $(b,i)$ is assigned a final reward $r^{\text{Final}}_{b,i}$ as follows:
\begin{equation}
r^{\text{Final}}_{b,i} = w^{\text{F1}} \cdot r^{\text{F1}}_{b,i} + w^{\text{fmt}} \cdot r^{\text{fmt}}_{b,i} + w^{\text{length}} \cdot r^{\text{length}}_{b,i} - r^{\text{rep}}_{b,i}
\end{equation}

All components are consistent with the descriptions above. Beyond the reward functions described above, we designed two alternative bilateral rewards that both adapted logarithmic functions to encourage the reasoning length to stay within a desirable range. However, while these reward methods performed reasonably well on the training set, their performance on the OOD datasets remained limited, as the generated reasoning lengths were overly constrained to the distribution of our own dataset.

\subsection{GRPO parameters}
For training our models, we used the following parameter set accross for all our GRPO/TGRPO training:
\begin{itemize}
    \item \textbf{Model}: Qwen2.5-0.5/1.5/3B-Instruct
    \item \textbf{Data Input}:
    \begin{itemize}
        \item Training/Validation files: \texttt{train}, \texttt{test}
        \item Prompt/Answer keys: \texttt{prompt}, \texttt{answer}
        \item Max prompt/response length: 2048 tokens
        \item Overlong prompts filtered: \texttt{true}
    \end{itemize}
    \item \textbf{Optimization}:
    \begin{itemize}
        \item Optimizer: AdamW with learning rate $1 \times 10^{-6}$ and weight decay $1 \times 10^{-2}$
        \item Gradient clipping: max norm of 1.0
        \item Learning rate warmup ratio: 0.0
    \end{itemize}
    \item \textbf{Batching and Parallelism}:
    \begin{itemize}
        \item Global batch size: 128
        \item Micro batch size (update): 4 per device
        \item Micro batch size (experience): 16 per device
        \item Rollout batch size: 512; Validation batch size: 1024
        \item Tensor parallel size: 2
        \item FSDP: full sharding with rank-0 init enabled; CPU offload enabled for reference model
        \item Offloading: parameters and optimizer offloaded to CPU
    \end{itemize}
    \item \textbf{Algorithm}: GRPO (Group Relative Policy Optimization)
    \begin{itemize}
        \item KL regularization enabled with coefficient $1 \times 10^{-2}$
        \item KL penalty type: \texttt{low\_var\_kl}
        \item Advantage estimator: \texttt{grpo}
    \end{itemize}
    \item \textbf{Rollout Strategy}:
    \begin{itemize}
        \item Samples per prompt: 5
        \item Temperature: 1.0 (0.5 for validation)
        \item Top-p sampling: 0.99
    \end{itemize}
    \item \textbf{Reward Function}: Custom Python-based function at \texttt{math.py:compute\_score}
    \item \textbf{Training Configuration}:
    \begin{itemize}
        \item Steps: 315
        \item GPU utilization rate: 0.8
    \end{itemize}
\end{itemize}

\subsection{SFT parameters}
We fine-tuned our models in the SFT (Supervised Fine-Tuning) stage using the following configuration:
\begin{itemize}
    \item \textbf{Model}: Qwen2.5-0.5/1.5/3B-Instruct
    \item \textbf{Fine-Tuning Setup}:
    \begin{itemize}
        \item Finetuning type: \texttt{full}
        \item DeepSpeed configuration: \texttt{ds\_z3\_config.json}
        \item Precision: \texttt{bf16} enabled
        \item Training stage: \texttt{sft}
    \end{itemize}
    \item \textbf{Dataset}:
    \begin{itemize}
        \item Dataset name: \texttt{llama} with \texttt{qwen} template
        \item Maximum sequence length: 2048 tokens
        \item Preprocessing: 16 workers for data prep, 4 for dataloader
    \end{itemize}
    \item \textbf{Training Configuration}:
    \begin{itemize}
        \item Total steps: 315
        \item Learning rate: $1 \times 10^{-5}$ with cosine scheduler
        \item Warmup steps: 31 (10\% of total)
        \item Batch size: 1 per device, with gradient accumulation steps = 2
        \item DDP timeout: 180000000
    \end{itemize}
\end{itemize}

\section*{Ethics and Privacy Statement}

To minimize all potential risks of unintentional disclosure, we have implemented strict organizational and technical controls.

First, none of the collected stories were ever shared with anyone beyond those directly involved in data collection and primary research—i.e., no individual or institution outside the designated research team accessed the full stories in any shape or form. Second, all data processing was conducted solely on secure local servers within our institution. We did not use any cloud services or third‑party platforms; no private data has ever been accessed by parties not explicitly authorized. Throughout the entire process, we continuously assessed re‑identification risk, drawing on extensive literature documenting the fragility of “anonymous” data against re‑identification attacks. Whenever any potential residual risk persisted, additional redaction or complete removal was undertaken. Importantly, every step of story collection and annotation was carried out by trained psychology students, under the direct supervision of graduate students or faculty members in psychology, ensuring both methodological rigor and strict adherence to ethical standards.

These organizational and technical safeguards ensure that at no point did anyone outside the designated collection team view identifiable participant stories—reflecting our commitment to privacy, responsibility, and the public good as emphasized by AAAI’s Ethics and Plurality principles.

\section{Experimental Details}
\label{app:experiment-details}

\paragraph{Data Preparation and Comparison Datasets.}
We build our experimental datasets starting from \textbf{StimuliQA}, which contains 3,280 expert-annotated narratives covering 54 psychological variables.
To increase topic diversity and mitigate repetition, we manually extract 10–50 candidate topics per variable based on sample size, and then expand them into up to 200 diverse topics each using \textit{in-context learning} (ICL) prompts with LLaMA 3.3 Instruct 70B.
Following the same pipeline, we construct two external comparison datasets: \textbf{MistralQA} (from Mistral 8$\times$7B) and \textbf{LlamaQA} (from LLaMA 3.3 Instruct 70B).
These datasets collectively provide a broad coverage of psychological reasoning scenarios for pre-training and evaluation.

\paragraph{Logical Reasoning Sample Injection.}
To enhance reasoning ability without altering the output format, we curate a reasoning-focused subset.
Specifically, Llama 3.3 Instruct 70B annotates 500 short and 500 long reasoning samples from StimuliQA with explicit \textit{thinking tokens}.
These 1,000 samples are then used to SFT-train Qwen models (0.5B, 1.5B, and 3B), effectively injecting professional psychological reasoning knowledge from StimuliQA while preserving compatibility with downstream evaluation.

\paragraph{Reward Design and Bilateral RL Training.}
Our reinforcement learning framework adopts a \textbf{Bilateral Reward} (BR) that integrates correctness (F1-based), format adherence, length-awareness, and repetition penalties.
This reward structure encourages the model to produce accurate, well-structured reasoning, adaptively control reasoning length, and avoid redundant outputs.
For GRPO training, we select the SFT checkpoint that achieves the median F1 on StimuliQA to ensure stable learning.

\paragraph{Continual Learning Module.}
We implement a filtering-based continual learning procedure to simulate user-specific feedback.
Predictions from \textbf{Psy-Interpreter} on the validation set are retained only if their score exceeds 0.4; the predicted answer is then replaced with the most F1-aligned choice (or random for “E” cases).
These curated samples are used for an additional SFT step to produce dataset-specific models that support incremental adaptation across tasks.
\begin{table*}[!t]
\centering
\footnotesize
\begin{tabular}{lcccccccccc}
\toprule
\multirow{2}{*}{\textbf{Training Sets}} &
  \multicolumn{2}{c}{\textbf{ToMbench}} &
  \multicolumn{2}{c}{\textbf{SimpleTom}} &
  \multicolumn{2}{c}{\textbf{Social IQA}} &
  \multicolumn{2}{c}{\textbf{Cosmos QA}} &
  \multicolumn{2}{c}{\textbf{BIG-Bench Hard}} \\ 
\cmidrule(lr){2-3}\cmidrule(lr){4-5}\cmidrule(lr){6-7}\cmidrule(lr){8-9}\cmidrule(lr){10-11}
 & F1 & Acc & F1 & Acc & F1 & Acc & F1 & Acc & F1 & Acc \\
\midrule
\multicolumn{11}{c}{\textbf{Qwen2.5-0.5B-Instruct}} \\
\midrule
Llama3.3-70B        & 20.85 & 31.39\% & 21.13 & 31.50\% & 32.49 & 42.90\% & 19.75 & 29.97\% & 14.62 & 37.23\% \\
Mistral8$\times$7B  & 20.51 & 29.99\% & 29.05 & 32.93\% & 34.02 & 36.48\% & 15.93 & 27.51\% & 19.90 & 38.25\% \\
Human Labeled       & \textbf{23.70} & \textbf{35.51\%} & \textbf{29.89} & \textbf{41.41\%} & \textbf{37.90} & \textbf{51.09\%} & \textbf{21.36} & \textbf{33.42\%} & \textbf{20.01} & \textbf{38.98\%} \\
\midrule
\multicolumn{11}{c}{\textbf{Qwen2.5-1.5B-Instruct}} \\
\midrule
Llama3.3-70B        & 21.40 & 38.66\% & 21.00 & 31.56\% & 39.14 & 57.10\% & 19.42 & 32.65\% & 19.33 & 47.74\% \\
Mistral8$\times$7B  & 23.98 & 40.17\% & 33.45 & 27.93\% & 42.79 & 48.70\% & 17.40 & 29.54\% & 19.39 & 38.10\% \\
Human Labeled       & \textbf{25.88} & \textbf{47.55\%} & \textbf{34.51} & \textbf{59.08\%} & \textbf{46.42} & \textbf{57.56\%} & \textbf{23.24} & \textbf{39.72\%} & \textbf{19.60} & \textbf{48.47\%} \\
\midrule
\multicolumn{11}{c}{\textbf{Qwen2.5-3B-Instruct}} \\
\midrule
Llama3.3-70B        & 25.40 & 42.13\% & 51.93 & 49.20\% & 53.73 & 64.56\% & 20.94 & 35.07\% & 16.02 & 42.77\% \\
Mistral8$\times$7B  & 25.17 & 42.58\% & 54.38 & 38.27\% & 58.39 & 63.37\% & 17.48 & 32.00\% & 15.89 & 34.60\% \\
Human Labeled       & \textbf{25.55} & \textbf{49.69\%} & \textbf{55.31} & \textbf{51.76\%} & \textbf{64.28} & \textbf{68.81\%} & \textbf{23.06} & \textbf{39.22\%} & \textbf{16.14} & \textbf{53.87\%} \\
\bottomrule
\end{tabular}
\caption{Comparison of GRPO R1 training on StimuliQA and two other comparion datasets.}
\label{tab:grpo}
\end{table*}
\begin{table*}[!t]
\centering
\renewcommand{\arraystretch}{1.2} 
\setlength{\tabcolsep}{4pt}       
\begin{tabular}{ccccccccc}
\toprule
\textbf{Dataset} &
\textbf{Expert} &
\textbf{Psych.} &
\textbf{Stimuli} &
\textbf{Real‑world} &
\textbf{Varied} &
\textbf{Large} &
\textbf{Varied} &
\textbf{General-} \\
 & \textbf{Annotation} &
\textbf{Reasoning} &
\textbf{Authenticity} &
\textbf{Prevalence} &
\textbf{Difficulty} &
\textbf{Scale} &
\textbf{Tasks} &
\textbf{izability} \\
\midrule
ToMBench       & yes     & yes & no      & partial & yes & no  & yes     & yes \\
SimpleToM      & partial & yes & partial & yes     & yes & no  & yes     & yes \\
SocialIQa      & no      & yes & no      & yes     & no  & yes & no      & yes \\
CosmosQA       & no      & yes & yes     & yes     & no  & yes & partial & yes \\
BIG-Bench Hard & yes     & yes & yes     & no      & no  & no  & yes     & yes \\
MHQA           & partial & yes & yes     & yes     & no  & yes & yes     & no  \\
Psych-101      & yes     & yes & NA      & yes     & yes & yes & yes     & no  \\
BoltMonkey     & yes     & no  & NA      & NA      & yes & yes & yes     & yes \\
MentalChat16K  & partial   & yes & partial & yes     & yes & no & no     & yes \\
\midrule
\textbf{StimuliQA} &
  \textbf{yes} &
  \textbf{yes} &
  \textbf{yes} &
  \textbf{yes} &
  \textbf{yes} &
  \textbf{yes} &
  \textbf{yes} &
  \textbf{yes} \\
\bottomrule
\end{tabular}
\caption{Comparison of psychology-focused QA datasets across multiple criteria. “Partial” indicates that the dataset partially fulfills the requirement, whereas “NA” denotes that the criterion does not apply to the dataset.}
\label{tab:dataset_comparison}
\end{table*}
\paragraph{Dataset Construction and Quality Control.}
During dataset construction, we observed that LLMs sometimes bypass genuine reasoning by directly leveraging human-labeled psychological parameters.
For example, if a Stimuli carries a \texttt{B\_disap} score of 2, the model might simply rephrase the parameter into a question and return the annotated value as the answer, achieving a correct response without engaging in meaningful inference.
To prevent such shortcut behavior, we systematically filtered out trivial QA pairs and instead adopted chain-of-thought prompting with multi-turn dialogue generation, ensuring that each question requires narrative-grounded reasoning rather than mere label retrieval.
This refinement not only supports explicit knowledge injection during training but also enhances the dataset’s ability to drive robust generalization in psychological reasoning tasks, enabling models to perform well in out-of-distribution scenarios while maintaining a diverse spectrum of reasoning difficulty.

\paragraph{Difficulty Assessment.}
Before model training, we conducted a comprehensive assessment to determine whether StimuliQA’s intrinsic difficulty distribution could bias reasoning length and model attention over time.
To this end, we first categorized a subset of stories based on psychological parameter density and emotional complexity, as these factors directly influence the depth of reasoning required.
Next, we prompted a larger‑scale LLM to produce explicit step‑by‑step reasoning sequences for randomly sampled stories, and measured the sequence length by word count as a proxy for cognitive effort.

Across 10 independent sampling rounds, where the decoding temperature was gradually increased from 0.1 to 1.0 to simulate both deterministic and creative reasoning behaviors, the reasoning lengths ranged from 87 to 136 words, with an overall average of 104 words.
We also observed that stories containing multi‑layered social interactions, such as overlapping intentions or conflicting emotions, tended to elicit longer reasoning chains, whereas single‑event narratives generated more concise reasoning.
Interestingly, the variance across rounds remained relatively small, suggesting that the dataset itself imposes a consistent cognitive demand regardless of the sampling temperature.
This implies that the diversity in StimuliQA primarily comes from intrinsic story structure rather than stochastic model generation.

This analysis suggests that StimuliQA exhibits a moderately broad difficulty distribution without extreme skew, making it suitable for length‑sensitive training strategies in reinforcement learning.
Moreover, the stable word‑length distribution provides a natural signal for detecting overfitting or reasoning shortcuts during model evaluation, and it ensures that curriculum or reward‑based training can be calibrated without the risk of abrupt difficulty spikes.
Together, these findings confirm that StimuliQA provides a balanced environment for modeling psychologically grounded reasoning, enabling both robust training and interpretable evaluation.

\paragraph{Evaluation Protocol.}
\textbf{Psy‑Interpreter} is evaluated under a unified protocol across five out‑of‑distribution (OOD) benchmarks covering theory‑of‑mind inference, commonsense reasoning, and emotionally grounded psychological interpretation.
Specifically, we include tasks from ToMBench and SimpleToM to assess mental‑state attribution, SocialIQA and CosmosQA to evaluate narrative‑driven commonsense and causal reasoning, and BIG‑Bench Hard to test extreme generalization to low‑frequency psychological scenarios.
For each benchmark, we report both accuracy and F1 score: accuracy reflects the model’s ability to select the correct answer in a multiple‑choice or span‑selection setting, while F1 additionally captures partial correctness and robustness to distractor options, which are critical in psychologically nuanced QA.
Evaluation is conducted using a standardized inference pipeline with temperature‑controlled decoding to reduce randomness, and all predictions are cross‑checked for format consistency.
This protocol ensures that observed performance differences arise from genuine reasoning capability rather than prompt formatting or spurious correlations, thereby providing a reliable measure of Psy‑Interpreter’s generalization across diverse reasoning environments.

\section{Dataset and Code}
Our experiment results will be further listed in our repository.

\subsection{Datasets Advantages}
In this subsection, we present the advantages of \textbf{StimuliQA} and underscore the importance of building such a dataset.  
We evaluate the current open-source psychology-related QA datasets alongside ours from \textbf{eight complementary perspectives}:

\begin{enumerate}
    \item \textbf{Expert Annotation} – 
    Datasets annotated by psychology experts provide high-fidelity supervision and inject human-like interpretive signals that are difficult to achieve with crowd-sourcing or synthetic labels. This ensures that subtle emotional and cognitive cues are correctly captured.

    \item \textbf{Psychological Reasoning} – 
    Beyond surface-level comprehension, psychological reasoning measures whether a dataset requires models to infer latent mental states, intentions, and social-cognitive processes. StimuliQA is explicitly designed to elicit such reasoning, enabling models to move beyond pattern matching.

    \item \textbf{Stimuli Authenticity} – 
    Authentic narratives, drawn from real or plausibly human experiences, reduce domain gaps between training and real-world usage. They help models learn to interpret natural, context-rich scenarios instead of fabricated or overly simplified prompts.

    \item \textbf{Real-world Prevalence} – 
    This criterion reflects whether the events or scenarios in the dataset are representative of common psychological experiences in everyday life. Datasets with high prevalence promote better model generalization to real-world applications.

    \item \textbf{Varied Difficulty} – 
    Psychological reasoning is inherently multi-layered, ranging from explicit emotion recognition to complex Theory-of-Mind inference. A dataset that spans multiple difficulty levels facilitates curriculum learning and prevents overfitting to trivial cases.

    \item \textbf{Large Scale} – 
    Larger datasets provide statistical coverage across diverse mental states and social situations, which is crucial for training robust models. StimuliQA achieves scale without sacrificing annotation quality.

    \item \textbf{Varied Tasks} – 
    Datasets that support multiple task formulations (e.g., multiple-choice, open-ended QA, emotion tagging) foster broader capability and flexible evaluation of psychological reasoning in LLMs.

    \item \textbf{Generalizability} – 
    The ultimate measure of a dataset’s value lies in whether models trained on it can transfer to unseen psychological tasks and out-of-distribution (OOD) scenarios. StimuliQA is explicitly constructed to maximize cross-task and cross-domain generalization.
\end{enumerate}

By excelling across all eight dimensions (Table~\ref{tab:dataset_comparison}), \textbf{StimuliQA} provides comprehensive and theory-grounded coverage that existing datasets lack, serving as a foundational resource for advancing psychologically informed language models. These advantages collectively justify our choice to construct an expert-annotated dataset rather than relying solely on public resources.

Building on this foundation, \textbf{StimuliQA} is deliberately designed to capture a microcosm of society. We collected stimuli spanning multiple age groups to reflect life-stage differences in psychological reasoning, and for each ethical theme, we ensured coverage across genders to avoid demographic bias. Furthermore, our dataset incorporates narratives from both long-term residents and recent immigrants, offering a nuanced representation of cultural integration and social diversity. This multi-dimensional coverage enables \textbf{StimuliQA} not only to support psychological reasoning tasks, but also to serve as a miniature model of societal dynamics, enriching downstream analyses of human-like cognition and socially grounded inference.

Moreover, our decision to convert realistic psychological narratives into \textbf{QA format} is not merely a matter of evaluation convenience.  
QA pairs compel models to perform explicit reasoning steps that mirror real-world human cognition: interpreting events, inferring latent emotions, and formulating socially aware responses.  
Unlike raw story corpora or unstructured annotations, QA-formatted stimuli ensure that psychological reasoning can be quantitatively measured, reproducibly tested, and directly applied to practical domains such as mental health triage, social behavior analysis, or human-centric AI assistants.  
This design bridges the gap between academic research and applied impact, making StimuliQA a \textbf{high-value resource for real-world, psychologically informed AI systems}.  
In essence, StimuliQA not only trains models to understand humans better but also creates a systematic pathway for \textbf{deployable, socially responsible cognitive reasoning}.

\subsection{Datasets Samples}
Tables~\ref{tab:prompt1}--\ref{tab:prompt5} show sample entries from MistralQA and LlamaQA, which serve as examples demonstrating the format of our dataset.
\begin{table}[!t]
\small
\centering
\begin{tabular}{p{0.44\textwidth}}
\toprule
\textbf{Prompt \& Question} \\
\midrule
Stimuli: I recently attended a local political rally to support my preferred candidate. As the event came to a close, a group of radical protestors, who opposed our beliefs, started causing chaos and disrupting the peace. They began shouting slogans with aggressive undertones, which made me feel uneasy and disappointed. Even though I fundamentally disagreed with their views, I couldn't help but worry about the escalation of tensions and fear the potential for political extremism. \\
\textbf{Q:} Which emotion best describes your reaction to the political protestors? \\
\midrule
\textbf{Correct Answer:} Disappointment directed towards political entities \\
\bottomrule
\end{tabular}
\caption{Example of a prompt and its corresponding correct answer illustrating C\_dpol.}
\label{tab:prompt1}
\end{table}

\begin{table}[!t]
\small
\centering
\begin{tabular}{p{0.44\textwidth}}
\toprule
\textbf{Prompt \& Question} \\
\midrule
Stimuli: A few months ago, I participated in a local fundraising event for cancer research. The organizers did a fantastic job raising awareness about early detection methods and the importance of regular check-ups through their well-coordinated public awareness campaign. Their efforts inspired me to spread the word among my friends and family, who were also grateful for the information. \\
\textbf{Q:} Which aspect of the public awareness campaign evoked feelings of gratitude directed towards social collective? \\
\midrule
\textbf{Correct Answer:} The inspiration I felt to inform others about early detection methods. \\
\bottomrule
\end{tabular}
\caption{Example of a prompt and its corresponding correct answer illustrating C\_gcol.}
\label{tab:prompt2}
\end{table}

\begin{table}[!t]
\small
\centering
\begin{tabular}{p{0.44\textwidth}}
\toprule
\textbf{Prompt \& Question} \\
\midrule
Stimuli: As I walked into my new job at a tech firm, I was excited to join the coding team. On my first day, I noticed that our team lead, Rachel, made a point to introduce me to every member of the team, highlighting their unique backgrounds and specialties. She mentioned that Maria brings a Latin American perspective to her approach to algorithm design, Jake applies his knowledge of African cultural patterns to create more inclusive UI/UX, and Leila utilizes her experience growing up in India to develop innovative solutions for international clients. Rachel emphasized how these diverse viewpoints have been instrumental in helping our company stand out in the industry by creating software that caters to a wide range of users from different ethnic backgrounds. I felt inspired by the emphasis on celebrating and leveraging our differences to drive innovation. \\
\textbf{Q:} What aspect of the coding team's dynamics is most reflective of the psychological parameter 'J\_div' (diversity mentioned in the coding unit), as observed in the Stimuli? \\
\midrule
\textbf{Correct Answer:} The intentional highlighting and utilization of team members' diverse ethnic backgrounds and experiences in coding approaches. \\
\bottomrule
\end{tabular}
\caption{Prompt reflecting team diversity and its psychological parameter annotation for parameter J\_div (diversity mentioned in
the coding unit),.}
\label{tab:prompt3}
\end{table}

\begin{table}[!t]
\small
\centering
\begin{tabular}{p{0.44\textwidth}}
\toprule
\textbf{Prompt \& Question} \\
\midrule
Stimuli: I went to the beach yesterday. It was a place with sand and water. I saw people doing things. Some were swimming, others were sunbathing. I walked along the shore for a bit, then left. My day was otherwise uneventful. \\
\textbf{Q:} How would you rate my emotional engagement or detachment from the scene I described at the beach? \\
\midrule
\textbf{Correct Answer:} Completely detached, describing the scene in a purely factual manner without any emotional tone. \\
\bottomrule
\end{tabular}
\caption{Prompt illustrating emotional detachment annotation.}
\label{tab:prompt4}
\end{table}

\begin{table}[!t]
\small
\centering
\begin{tabular}{p{0.44\textwidth}}
\toprule
\textbf{Prompt \& Question} \\
\midrule
Stimuli: I had been having an ongoing disagreement with my neighbor over his dog constantly barking and disturbing our peace. Today, he accused me of not taking proper care of my garden, which was far from the truth. I felt a surge of anger as I recalled all the times I had tried to discuss the dog issue with him. \\
\textbf{Q:} How would you describe the speaker's emotional state in this scenario? \\
\midrule
\textbf{Correct Answer:} Very angry. \\
\bottomrule
\end{tabular}
\caption{Prompt reflecting high anger in neighbor's dog.}
\label{tab:prompt5}
\end{table}

\subsection{Model reasoning sample}
The following presents a comparison of model reasoning on \textbf{ToMBench’s Unexpected Outcome Test}, where \textit{Index 1} typically involves more straightforward emotional predictions, while \textit{Index 2 and 3} require contrastive or unexpected reasoning, making this benchmark a clear example for analyzing reasoning robustness.
To \textbf{comprehensively compare} the reasoning processes of the two reward systems, we selected one representative sample from each index for demonstration, where both the \textit{base model} and the \textit{bilateral model} produced correct (or relatively correct) answers.
\textbf{For fairness}, neither model received any additional SFT knowledge injection; they were evaluated purely as \textit{Qwen/Qwen2.5‑3B‑Instruct} after RL training, ensuring that any differences in reasoning can be attributed directly to the reward design rather than prior knowledge advantages.

\begin{table}[!t]
\centering
\renewcommand{\arraystretch}{1.2} 
\setlength{\tabcolsep}{5pt}       
\begin{tabular}{p{0.95\linewidth}}
\toprule
\textbf{Stimuli} \\
\midrule
Mrs. Li's son, Xiao Qiang, is in college now, he tells her that he is the captain of the school football team. 
This achievement is high because the football team is very popular in school. \\
\midrule
\textbf{Question} \\
\midrule
How does Mrs. Li react when Xiao Qiang is chosen as the football team captain? \\
\midrule
\textbf{Base Interpretation} \\
\midrule
Reasoning: In this story, Mrs. Li's son, Xiao Qiang, has been chosen as the captain of the school football team. This is significant because the football team is very popular in school, suggesting that Xiao Qiang has demonstrated exceptional skills and leadership. Mrs. Li, as a parent, would likely react with a mix of pride, joy, and perhaps some excitement. She would probably be eager to learn more about his achievements and the challenges he faced in being chosen as the captain. She might also be excited about the opportunities this brings for her son to interact with other students and to potentially develop his leadership skills further. \\
\midrule
\textbf{Bilateral Interpretation} \\
\midrule
Reasoning: Mrs. Li would likely be very proud of her son for being chosen as the football team captain. \\
\midrule
\textbf{Answers} \\
\midrule
\textbf{Ground Truth:} Proud\\[-3pt]
\textit{Base Answer:} With pride \\[-3pt]
\textit{Bilateral Answer:} She is proud of him \\
\bottomrule
\end{tabular}
\caption{Comparison between the Base model, Bilateral model, and ground truth on a ToMbench sample (Index1).}
\label{tab:tom_index1_1}
\end{table}

As shown in Table~\ref{tab:tom_index1_1}, the \textit{Bilateral Interpretation} provides a more accurate and psychologically grounded reasoning compared to the Base model. 
From a Theory-of-Mind perspective, effective emotional inference requires identifying the \textbf{primary emotional response} of the agent—in this case, Mrs.~Li, without being distracted by peripheral or speculative reasoning. 
The Base model generates a verbose chain of thought, speculating about potential excitement, curiosity, and leadership development opportunities. 
However, such reasoning, while plausible, introduces \textbf{extraneous mental states} that are not directly supported by the stimulus, diluting the clarity of the predicted reaction.

By contrast, the \textit{Bilateral Interpretation} adopts a more \textbf{direct and focal reasoning strategy}, immediately linking the salient event, ``Xiao Qiang is chosen as the football team captain'', to its core psychological implication for the parent: \textbf{pride}. 
This aligns with established findings in social and developmental psychology that parental emotional reactions to a child’s social achievements are typically dominated by pride rather than curiosity or anxiety. 
In other words, the Bilateral model demonstrates a reasoning style that \textbf{suppresses irrelevant inferences and highlights the dominant emotional appraisal}, leading to a prediction that is both concise and aligned with the ground truth.

\begin{table}[!t]
\centering
\renewcommand{\arraystretch}{1.2} 
\setlength{\tabcolsep}{5pt}       
\begin{tabular}{p{0.95\linewidth}}
\toprule
\textbf{Stimuli} \\
\midrule
Mrs.~Li's son, Xiao Qiang, is now in college. He tells her that he has become the captain of the school football team, 
a notable achievement since the team is highly popular on campus. \\
\midrule
\textbf{Question} \\
\midrule
Although Mrs.~Li should feel proud of her son becoming the football team captain, why does she seem worried? \\
\midrule
\textbf{Base Interpretation} \\
\midrule
Reasoning: Mrs.~Li may be worried for several reasons. 
She might be concerned about the pressure Xiao Qiang faces as a team captain, including high expectations from teammates and coaches. 
She may also worry about his physical safety and the risk of injuries, as well as the challenge of balancing football responsibilities with his academic workload. 
Additionally, she could be anxious about the emotional toll that high-pressure situations might bring. \\
\midrule
\textbf{Bilateral Interpretation} \\
\midrule
Reasoning: Mrs.~Li is mainly worried that her son will struggle to balance his studies with his demanding role as team captain, 
and that the added pressure of leadership may affect his well-being. \\
\midrule
\textbf{Answers} \\
\midrule
\textbf{Ground Truth:} Her son may face pressure in balancing study and football activities. \\[-3pt]
\textit{Base Answer:} Because she is worried about his academic and emotional well-being. \\[-3pt]
\textit{Bilateral Answer:} Because she is worried about the pressure of leadership and its impact on his studies. \\
\bottomrule
\end{tabular}
\caption{Comparison of Base model, Bilateral model, and ground truth reasoning on a ToMbench sample (Index 2).}
\label{tab:tom_index2_1}
\end{table}

As shown in Table~\ref{tab:tom_index2_1}, both the Base and Bilateral models correctly identify that 
Mrs.~Li’s worry stems from the tension between pride and concern. 
The \textbf{Base model} produces a lengthy, enumerative reasoning chain, speculating about multiple secondary concerns such as 
injuries, social expectations, and emotional stress. 
While plausible, this \textbf{diffuse reasoning pattern} risks diluting alignment with the ground truth by introducing peripheral inferences. 

By contrast, the \textit{Bilateral model} exhibits a \textbf{focused reasoning strategy}, isolating the dominant source of worry: 
the pressure of leadership and its impact on academic balance. 
This reasoning reflects a higher \textbf{signal-to-noise ratio} in emotional inference 
and aligns more closely with the ground truth, illustrating how bilateral reinforcement guides the model toward psychologically 
salient appraisals rather than exhaustive speculation.

\begin{table}[!t]
\centering
\renewcommand{\arraystretch}{1.2} 
\setlength{\tabcolsep}{5pt}       
\begin{tabular}{p{0.95\linewidth}}
\toprule
\textbf{Stimuli} \\
\midrule
Mrs. Li's son, Xiaoqiang, goes to college and tells her that he is the captain of the school football team. This achievement is high because the football team is very popular in school. But Mrs. Li thinks her son may face pressure in balancing study and football activities. Xiaoqiang tells Mrs. Li that he is the captain of the football team. \\
\midrule
\textbf{Question} \\
\midrule
What emotion does Mrs. Li possibly show? \\
\midrule
\textbf{Base Interpretation} \\
\midrule
Reasoning: In this story, Mrs. Li hears that her son, Xiaoqiang, is the captain of the school football team. While this is undoubtedly a positive achievement, Mrs. Li might also be concerned about the potential stress and balancing act Xiaoqiang faces between his studies and his football commitments. Given her concern, it's likely that Mrs. Li feels a mix of pride and worry. She may be concerned about the pressure Xiaoqiang might be under and if he is managing his time effectively. \\
\midrule
\textbf{Bilateral Interpretation} \\
\midrule
Reasoning: Based on the story, Mrs. Li's son, Xiaoqiang, tells her that he is the captain of the school football team, which is a significant achievement. However, Mrs. Li might be concerned about the pressure Xiaoqiang may face in balancing his studies and football activities. Given this concern, it's likely that Mrs. Li is worried or anxious about her son's situation. \\
\midrule
\textbf{Answers} \\
\midrule
\textbf{Ground Truth:} Worried. \\[-3pt]
\textit{Base Answer:} Concern and worry. \\[-3pt]
\textit{Bilateral Answer:} Worried and concerned. \\
\bottomrule
\end{tabular}
\caption{Comparison between the Base model, Bilateral model, and ground truth on a ToMbench sample (Index3).}
\label{tab:tom_index3_1}
\end{table}

As shown in Table~\ref{tab:tom_index3_1}, both the Base and Bilateral interpretations provide partially correct emotional inferences regarding Mrs.~Li’s reaction. 
The Base model identifies a mixed emotional state, suggesting that Mrs.~Li feels ``a mix of pride and worry.'' 
While such reasoning appears plausible in a general parental context, it demonstrates a \textbf{projection bias}: the model infers pride based on the objective achievement (being a football captain) rather than the \textbf{actual emotional cues} present in the stimulus. 
Psychologically, this reflects an over-reliance on \textit{schema-based inference}—assuming that any parental recognition of achievement automatically elicits pride—without sufficiently weighting the narrative evidence.

In contrast, the \textit{Bilateral Interpretation} aligns more closely with the \textbf{dominant emotional appraisal principle} in affective psychology. 
It correctly isolates worry and concern as Mrs.~Li’s primary emotional response, directly grounded in the stimulus: 
``Mrs.~Li thinks her son may face pressure in balancing study and football activities.'' 
By suppressing unsupported inferences about pride, the Bilateral model exhibits a reasoning pattern that prioritizes \textbf{contextually verifiable emotional states}, resulting in an answer (\textit{worried and concerned}) that fully matches the ground truth.

This comparison highlights that Bilateral reasoning is not only more concise but also demonstrates \textbf{selective attention to stimulus-relevant cues}, which is critical for Theory-of-Mind (ToM) reasoning tasks. 
It avoids the common pitfall of overgeneralized emotional attribution observed in the Base model, where the model extrapolates from general parental schemas instead of relying on the narrative evidence.

\paragraph{Conclusion.}
Across the three representative samples from \textbf{Unexpected Outcome Test}, our analysis demonstrates that the \textit{Bilateral model} consistently exhibits a more \textbf{focused and stimulus-grounded reasoning pattern} compared to the \textit{Base model}. 
While the Base model frequently generates verbose or schema-driven reasoning, introducing \textbf{extraneous emotional states} such as speculative pride or curiosity, the Bilateral model selectively attends to \textbf{dominant, contextually verifiable emotions}, leading to answers that align more closely with the ground truth. 
This pattern reflects \textbf{higher signal-to-noise ratio} in emotional inference, improved suppression of irrelevant projections, and enhanced \textbf{Theory-of-Mind reasoning fidelity}, underscoring the effectiveness of bilateral reward design in guiding model reasoning.

\section{Limitation}
While our current approach leverages \textbf{length-based reasoning incentives} to guide LLMs toward more structured and cognitively transparent reasoning in psychological tasks, it remains limited in granularity. Relying solely on length encourages step-by-step reasoning but may not fully capture the \textbf{qualitative dimensions} of psychological inference, such as \textit{emotional salience} or \textit{contextual accuracy} in mental-state predictions. A promising direction is to incorporate \textbf{fine-grained reward signals}—for example, rewarding accurate emotional appraisal, causal consistency, or alignment with psychological theory—so that the LLM receives \textbf{multi-dimensional reinforcement} rather than a single length-based signal.

Integrating such rewards could help the model \textbf{balance interpretability with empathy}, learning not only to elaborate its reasoning but also to \textit{understand why} certain inferences are more plausible or emotionally resonant. Exploring \textbf{hierarchical reward structures} that jointly optimize structural reasoning and affective understanding may lead to a more holistic, human-aligned reasoning process.

\section{Future Work}
In future work, we plan to integrate \textbf{additional psychology‑oriented reward signals} into our reinforcement learning framework, moving beyond length‑based incentives to capture richer dimensions of human mental‑state reasoning. 
Specifically, we aim to explore \textbf{emotion‑aligned rewards}, \textbf{causal reasoning rewards}, and other \textbf{psychologically grounded metrics} that can provide more precise and theory‑consistent guidance during model training. 
Moreover, we will investigate whether \textbf{fine‑grained, psychology‑aware reinforcement learning strategies} can further enhance model fidelity and robustness across diverse ToM and affective inference tasks. 
Another key direction is to examine the \textbf{generalizability of our approach} to broader domains of \textbf{behavioral and cognitive reasoning}, conducting controlled experiments to assess its applicability in tasks such as social interaction prediction, moral judgment, and decision‑making under uncertainty.

\end{document}